\documentclass[useAMS,usenatbib]{mn2e}

\usepackage{graphicx}	
\usepackage{amsmath}	
\usepackage{amssymb}	
\usepackage{multicol}        
\usepackage{bm}		
\usepackage{pdflscape}	
\usepackage{multirow,multicol} 

\def\N1316{NGC\,1316}
\def\N1404{NGC\,1404}

\def\4U{4U~1735$-$444}

\def\arcsec{\ifmmode '' \else $''$\fi}

\def\arcsecpoint{\ifmmode ''\!. \else $''\!.$\fi}

\def\kms{\ifmmode {\rm km\ s}^{-1} \else km s$^{-1}$\fi}
\def\Msun{\ifmmode {\rm M}_{\odot} \else M$_{\odot}$\fi}
\def\Lsun{\ifmmode {\rm L}_{\odot} \else L$_{\odot}$\fi}
\def\Zsun{\ifmmode {\rm Z}_{\odot} \else Z$_{\odot}$\fi}

\def\ergscm2{ergs\,s$^{-1}$\,cm$^{-2}$}
\def\icm3{{\rm cm}^{-3}}
\def\icm2{{\rm cm}^{-2}}
\def\qo{\ifmmode q_{\rm o} \else $q_{\rm o}$\fi}
\def\Ho{\ifmmode H_{\rm o} \else $H_{\rm o}$\fi}
\def\ho{\ifmmode h_{\rm o} \else $h_{\rm o}$\fi}

\def\vFWHM{\ifmmode v_{\mbox{\tiny FWHM}} \else
            $v_{\mbox{\tiny FWHM}}$\fi}
\def\CCF{\ifmmode F_{\it CCF} \else $F_{\it CCF}$\fi}
\def\ACF{\ifmmode F_{\it ACF} \else $F_{\it ACF}$\fi}
\def\Halpha{\ifmmode {\rm H}\alpha \else H$\alpha$\fi}
\def\Hbeta{\ifmmode {\rm H}\beta \else H$\beta$\fi}
\def\Hgamma{\ifmmode {\rm H}\gamma \else H$\gamma$\fi}
\def\Hdelta{\ifmmode {\rm H}\delta \else H$\delta$\fi}
\def\Lya{\ifmmode {\rm Ly}\alpha \else Ly$\alpha$\fi}
\def\Lyb{\ifmmode {\rm Ly}\beta \else Ly$\beta$\fi}
\def\Lyg{\ifmmode {\rm Ly}\beta \else Ly$\gamma$\fi}

\def\ciii{\ifmmode {\rm C}\,{\sc iii} \else C\,{\sc iii}\fi}
\def\civ{\ifmmode {\rm C}\,{\sc iv} \else C\,{\sc iv}\fi}
\def\cv{\ifmmode {\rm C}\,{\sc v} \else C\,{\sc v}\fi}
\def\cvi{\ifmmode {\rm C}\,{\sc vi} \else C\,{\sc vi}\fi}

\def\o5007{[O\,{\sc iii}]\,$\lambda5007$}

\def\oviiviii{O\,{\sc vii-viii}}

\def\fexxii-iii{Fe\,{\sc xxii-xxiii}}

\title[UFOs disappear in high radiation fields]{Ultrafast outflows disappear in high radiation fields}
\author[C. Pinto et al.]{C. Pinto,$^{1}$\thanks{E-mail:
cpinto@ast.cam.ac.uk} W. Alston,$^{1}$ M. L. Parker,$^{1}$ A. C. Fabian,$^{1}$  
\newauthor L. C. Gallo,$^{2}$ D. J. K. Buisson,$^{1}$ D. J. Walton,$^{1}$ E. Kara,$^{3}$ 
\newauthor J. Jiang,$^{1}$ A. Lohfink,$^{1,4}$ and C. S. Reynolds\,$^{3,5}$ \\  
$^{1}$ Institute of Astronomy, Madingley Road, CB3 0HA Cambridge, United Kingdom\\
$^{2}$ Department of Astronomy and Physics, Saint Mary's University, 923 Robie Street, Halifax, NS B3H 3C3, Canada\\
$^{3}$ Department of Astronomy, University of Maryland, College Park, MD 20742-2421, USA\\
$^{4}$ Department of Physics, Montana State University, P.O. Box 173840, Bozeman, MT 59717-3840, USA\\
$^{5}$ Joint Space Science Institute, University of Maryland, College Park, MD 20742, USA}
\begin{document}

\date{\today}

\pagerange{\pageref{firstpage}--\pageref{lastpage}} \pubyear{2017}

\maketitle

\label{firstpage}

\begin{abstract}
Ultrafast outflows (UFOs) are the most extreme winds launched by 
active galactic nuclei (AGN) due to their mildly-relativistic speeds ($\sim0.1-0.3c$) 
and are thought to significantly contribute to galactic evolution via AGN feedback.
Their nature and launching mechanism are however not well understood. 
Recently, we have discovered the presence of a variable UFO in
the narrow-line Seyfert 1 IRAS 13224$-$3809.
The UFO varies in response to the brightness of the source. 
In this work we perform flux-resolved 
X-ray spectroscopy to study the variability of the UFO and found that the ionisation parameter
is correlated with the luminosity. In the brightest states the gas is almost completely ionised
by the powerful radiation field and the UFO is hardly detected. 
This agrees with our recent results obtained with principal component analysis.
We might have found the tip of the iceberg:
the high ionisation of the outflowing gas may explain why it is commonly difficult 
to detect UFOs in AGN and possibly suggest that we may underestimate their actual feedback. 
We have also found a {tentative} correlation between the outflow velocity and the luminosity,
which is expected from theoretical predictions of radiation-pressure driven winds.
{This trend is rather marginal due to the Fe {\small XXV-XXVI} degeneracy. 
Further work is needed to break such degeneracy through time-resolved spectroscopy}.
\end{abstract}



\section{Introduction}
\label{sec:intro}

Active galactic nuclei (AGN) are powered by accretion onto supermassive
black holes (SMBH) with conversion of gravitational energy into vast amounts of radiation
emitted over the entire electromagnetic spectrum. 
The energy throughput of AGN can regulate the growth of their host galaxies through
the process known as AGN feedback (e.g. Fabian 2012 and references therein). 
Gas outflows in the form of winds and powerful jets release a significant amount of energy 
and momentum into the interstellar medium (ISM) which can expel the gas otherwise 
used to generate new stars. 
Alternatively, AGN outflows may trigger star formation by compressing the gas
within the process known as
positive feedback (see e.g. Maiolino et al. 2017).
Ultrafast outflows (UFOs, e.g. Cappi et al. 2009) are the most extreme among the 
AGN winds for their mildly-relativistic speeds ($\gtrsim10,000$ km/s or $0.03c$) 
and amount of energy ($\gtrsim 0.05 L_{\rm Edd}$). 
These extreme winds are believed to originate from the accretion disc within a hundred
gravitational radii from the black hole. 

UFOs are commonly detected through high-excitation Fe {\scriptsize XXV-XXVI} absorption lines 
in the hard X-ray band ($7-10$\,keV, see e.g. Tombesi et al. 2010 and references therein). 
This energy band is very useful because of the much lower confusion with absorption 
from low-velocity warm absorbers commonly found in Seyfert galaxies. 
Warm absorbers imprint strong absorption
features in the soft X-ray band due to their lower ionisation parameter, 
but their contribution to feedback
is small (see e.g. Krongold et al. 2007 and references therein).
Moreover, the Fe {\scriptsize XXV-XXVI} lines are the strongest absorption lines produced 
at high ionisation, which is expected for winds originating in the innermost 
hot regions of the accretion discs.
However, the presence of reflection features (e.g. reprocessing of the primary emission from
the accretion disc) may complicate the identification of UFO Fe K absorption lines
(see e.g. Gallo and Fabian 2011).

The soft X-ray band ($\sim 0.3-2$\,keV) also provides a useful tool to search for extreme winds, 
but care has to be taken in distinguishing UFO features from absorption lines produced by
the Galactic interstellar medium (ISM, e.g. Pinto et al. 2013) and warm absorbers present near 
the X-ray source (e.g. Porquet and Dubau 2000).
There are recent UFO detections in ultraluminous X-ray sources 
in the soft band with the high-resolution reflection grating spectrometers (RGS) on board 
\textit{XMM-Newton} (Pinto et al. 2016 and Kosec et al. 2017)
as well as several claims in AGN high-resolution spectra 
{(e.g. Gupta et al. 2013, 2015, Pounds 2014, Longinotti et al. 2015, and Reeves et al. 2016)}.
These detectors and the gratings on board Chandra provide the highest spectral resolution 
and resolving power in the X-ray band, which is very useful to detect and study UFOs.

UFOs show clear signatures of variability in multi-epoch deep observations
with a possible transient nature (e.g. Dadina et al. 2005, Dauser et al. 2012, 
Zoghbi et al. 2015). The exact nature of their variability and their launching 
mechanism are not well understood. 
Variations in the column density of the absorber have been invoked to 
describe the variability of the wind features in PDS 456 and 1H 0707--495
(see, e.g., Reeves et al. 2009 and Hagino et al. 2016).
In a recent very deep campaign (PI: Fabian) on the 
narrow-line Seyfert 1 (NLS1) IRAS 13224$-$3809, hereafter IRAS 13224, 
we have discovered a variable ultrafast outflow  
with a clear anti-correlation between the strength of the absorption features 
(both in the soft and the hard X-ray bands) and the luminosity (Parker et al. 2017a, hereafter P17a). 
This was also confirmed by follow up work using principal component analysis (PCA, 
Parker et al. 2017b, hereafter P17b). Our results argued in favour of a change in the
ionisation state of the outflowing gas with values sufficiently high in the brightest states that 
the gas is almost completely ionised and any absorption features are significantly weakened.

\begin{figure*}
  \includegraphics[width=2.0\columnwidth]{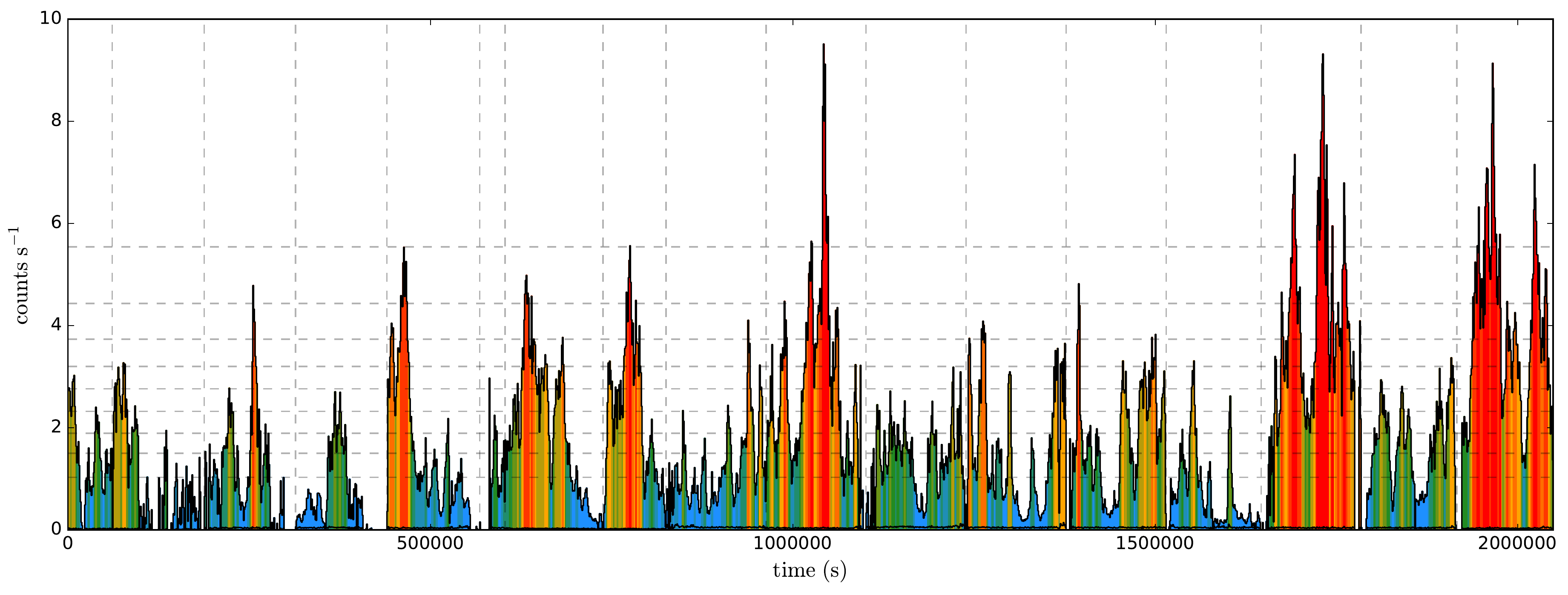}
   \caption{EPIC/pn lightcurve of all the observations in the 0.3--10.0\,keV energy range.  
            The vertical lines indicate the start and the end of each observation.
            The horizontal lines and the colours show the thresholds of the flux intervals.
            Flux thresholds for 10 flux regimes were chosen in order have comparable statistics.
            Time gaps between the observations are not shown.
            For more detail see text and Table\,\ref{table:fluxes}.} 
            \label{Fig:Fig_lightcurve}
   \vspace{-0.5cm}         
\end{figure*}

 \begin{table}
 \caption{{EPIC-pn} flux levels selection and exposure times.}  
 \label{table:fluxes}      
 \renewcommand{\arraystretch}{1.1}
  \small\addtolength{\tabcolsep}{0pt}
  
 \scalebox{1}{%
 \begin{tabular}{c c c c c c c c c c}     
 \hline  
 Level       &  Minimum     & Maximum     & Frame    & pn      & RGS  \\
                 &     (c/s)         &       (c/s)        & (ks)        & (ks)    & (ks)   \\
 \hline                                                                                                
1               &  $-$               &  1.135          &  5.04     &  443   &  572   \\
2               & 1.135            &  1.613          &  1.770   &  197   &  232   \\
3               & 1.613            &  2.023          &  1.543   &  148   &  196   \\
4               & 2.023            &  2.484          &  1.478   &  120   &  157   \\
5               & 2.484            &  2.895          &  1.386   &   97    &  118    \\
6               & 2.895            &  3.337          &  1.421   &   93    &  118    \\
7               & 3.337            &  3.884          &  1.375   &   69    &  101    \\
8               & 3.884            &  4.605          &  1.470   &   67    &   81    \\
9               & 4.605            &  5.761          &  1.933   &   53    &   75    \\
10             & 5.761            &  $-$              &  3.0       &   37    &   42    \\
 \hline                  
 \end{tabular}}  
    
Notes: the frame duration refers to the average length
of each section in the EPIC-pn lightcurve for a given flux interval; 
the total exposure is the time integration for each flux interval along 
all archival and new pn/RGS observations. 
   \vspace{-0.5cm}         
 \end{table}

In this work we have performed flux-resolved X-ray spectroscopy to study the variability 
of the UFO and confirmed that its variability can be explained in terms of increase in the 
ionisation parameter with the luminosity. 
This was possible through the combination of the high effective area of the CCD detectors
and the high spectral resolution of the gratings on board \textit{XMM-Newton}.

This paper is structured as follows.  
In Sec.\,\ref{sec:source} we report some well known characteristics of IRAS 13224
that motivate us to search for evidence of winds. We present the data in Sec.\,\ref{sec:data}
and a detailed spectral modelling in Sec.\,\ref{sec:spectral_modelling}.
We discuss the results and provide some insights on future missions in Sec.\,\ref{sec:discussion} 
and give our conclusions in Sec.\,\ref{sec:conclusion}.

\section[]{IRAS 13224$-$3809}
\label{sec:source}

IRAS 13224 is a low-redshift ($z=0.0658$) NLS1 galaxy. 
Such objects typically exhibit high mass accretion rates
onto black holes with masses smaller than quasars 
($10^{6-7} M_{\odot}$, see, e.g., Jin et al. 2012 and Chiang et al. 2015). 
The smaller masses significantly shorten the timescales of the variability processes 
and facilitate detailed study with the use of a few nearby exposures. 
The same analysis for quasars would require several years 
of observations. Among the brightest nearby NLS1, IRAS 13224 has the highest X-ray variability
(see e.g. Ponti et al. 2012), 
with luminosity jumps of two orders magnitude in the X-ray band occurring on very short time frames 
of just a few hours (see Fig.\,\ref{Fig:Fig_lightcurve}). 
We observed this source with a deep, 1.5\,Ms, \textit{XMM-Newton} campaign (PI: Fabian) 
to study its variability in great detail, particularly for its strong relativistic reflection (high-energy photons 
reprocessed by the inner accretion disc) and time lags between the power-law continuum and 
the reflection features. Additional observations totalling 500ks are available in the \textit{XMM-Newton}
archive\footnote{https://www.cosmos.esa.int/web/xmm-newton/xsa}.

The large amount of data allowed us to detect multiple absorption features produced by a $\sim0.2c$ 
UFO totalling a significance above $7\,\sigma$ (P17a, P17b). We also found that the strength of the 
absorption features decreased with increasing luminosity. 
Owing to the extreme flux and absorption variability 
present in IRAS 13224, this dataset provides us with the best framework to study the link
between the source luminosity and the UFO (and its detection).

\section[]{The data}
\label{sec:data}

In this paper we use all \textit{XMM-Newton} observations of IRAS 13224 
including the $\sim500$\,ks archival and $\sim1.5$\,Ms new data. 
We utilize data from the broadband {($0.3-10$\,keV)} EPIC-pn CCD spectrometer 
(Turner et al. 2001) and the high-resolution Reflection Grating Spectrometer 
(RGS, {$0.35-1.77$\,keV}, den Herder et al. 2001) in order to constrain the spectral shape of the 
source and the main characteristics of the wind with a focus on their variability.

The data reduction follows the same steps used in P17a. 
Briefly, we reduced the data with the latest \textit{XMM-Newton} Science Analysis System
(SAS) v15.0.0 (CALDB available on April, 2017) and the standard SAS threads. 
EPIC-pn data were reduced with the {\scriptsize EPPROC} 
task and corrected for contamination from soft-proton flares.

In order to study the variability of the UFO with the source luminosity, we split the entire dataset 
into 10 different flux levels, such that the total number of counts in each of the 10 spectra is comparable.
We pursued this by extracting a lightcurve for each observation and merging them into an overall
lightcurve which is then divided into 10 flux intervals 
(see Fig.\,\ref{Fig:Fig_lightcurve} and Table\,\ref{table:fluxes}). 
We used the standard threads of the
\textit{XMM-Newton}/SAS\footnote{https://www.cosmos.esa.int/web/xmm-newton/sas} software.

For each flux interval, 
we extracted EPIC-pn spectra from within a circular region of 1 arcmin diameter centred on the 
emission peak. The background spectra were extracted from within a larger circle in a nearby
region on the same chip, but away from the readout direction and the high copper background 
(Lumb et al. 2002). We also tested three other background regions on different chips to confirm that 
the background does not produce instrumental features.

The RGS data reduction was performed with {\scriptsize RGSPROC}.
We extracted the first-order RGS spectra in a cross-dispersion region of 1 arcmin width, 
centred on the emission peak and extracted the background spectra by selecting photons
beyond the 98 per cent of the source point-spread-function.
For completeness, we checked that the background spectra were consistent
with blank field observations. 

We stacked EPIC-pn and RGS 1--2 spectra from time intervals with same flux level
and obtain 20 final spectra (one EPIC-pn and one RGS for each flux level). 
Each pn flux-resolved spectrum has approximately 290k counts, 
whilst its corresponding RGS spectrum has 33k counts. 
The exposure times for all spectra are shown in Table\,\ref{table:fluxes}.
We note that the RGS exposure is actually double in the wavelength ranges common
to the RGS 1 and 2 detectors ($5.2-10.6$\,{\AA}, $13.8-20.0$\,{\AA}, $24.1-37.3$\,{\AA});
RGS 1 misses the $10.6-13.8$\,{\AA} chip, whilst RGS 2 misses the $20.0-24.1$\,{\AA}
chip. {The RGS exposure time for each flux interval accounts for the chip losses
and the stacking of RGS 1 and 2. This explains the different exposure times
in Table \ref{table:fluxes}.}

We do not use data from the MOS 1 and 2 cameras because each has 3--4 times less 
effective area (and even less above 8\,keV) than pn, 
which alone contains the vast majority of the counts.
We grouped the pn spectra in bins with {a signal-to-noise ratio of at least 6} and use $\chi^2$ statistics.
Throughout the paper we adopt 1\,$\sigma$ error bars.
We perform all spectral fits with the new {\scriptsize{SPEX}} code{\footnote{http://www.sron.nl/spex}} v3.03. 

The EPIC-pn and RGS flux-resolved spectra are shown in Fig.\,\ref{Fig:Fig_pn_spectra}
and \ref{Fig:Fig_rgs_spectra}, respectively.
The main signature of the UFO in the RGS spectrum is a broad O {\scriptsize VIII} 
absorption feature at around 16\,{\AA} together with some 
other weaker features between 10 and 13\,{\AA},
which are much more significant in deeper spectra extracted with fewer flux levels (see P17a).
In the EPIC-pn spectra, the UFO can be recognised by a broad absorption feature around 8\,keV 
plus other weaker features between 2 and 5\,keV due to high-excitation states
of magnesium, silicon, sulphur, argon and calcium (P17b).

\section{Flux-resolved X-ray spectroscopy}
\label{sec:spectral_modelling}

In P17a, using the same flux levels, we found that the equivalent
width of the Fe K absorption feature decreases with the source flux. 
It is therefore a useful exercise to apply self-consistent
physical models of absorption from gas in photoionised equilibrium to understand which parameters
mainly vary in the wind when the luminosity increases.

\vspace{-0.5cm}

\subsection{Broadband EPIC spectral continuum}
\label{sec:epic_continuum}

The exact knowledge of the continuum spectral components is beyond the aim of this paper
and will be widely discussed in a companion paper (Jiang et al. submitted).
Therefore, we adopted a phenomenological model to describe the broadband spectral
continuum, which consists of a power law (primary continuum) with slope 
$\Gamma \sim 2.2-2.9$, a blackbody (soft excess) with temperature $k T \sim 0.11-0.12$ keV
and two Gaussian lines relativistically-broadened (i.e. multiplied by the {\scriptsize LAOR} model 
in {\scriptsize{SPEX}} with emissivity index of about 9, 
inclination $\sim$ 70 degrees, and inner radius 1.4$r_g$)
to describe the main reflection features around 1 and 7 keV following
the approach used by Fabian et al. (2013). 
The emission model is corrected for redshift and absorption due to the foreground 
interstellar medium ({\scriptsize {HOT}} model in {\scriptsize{SPEX}} with column density 
$N_{\rm H}=6.75 \times 10^{20} {\rm cm}^{-2}$ 
and low temperature $0.5$\,eV, see e.g. Pinto et al. 2013). 
We adopted the column density from the Willingale 
tool\footnote{http://www.swift.ac.uk/analysis/nhtot/} 
to account for any contribution from both gas and molecules. 
The detailed spectral fits for the continuum of the 10 flux-resolved spectra are given 
in Appendix \ref{sec:appendix} and Table \ref{table:table_continuum}.
The 10 EPIC-pn spectra are fitted simultaneously in order to better constrain
the parameters of the relativistic broadening ({\scriptsize LAOR} model).

\begin{figure}
  \includegraphics[width=1.\columnwidth,angle=0]{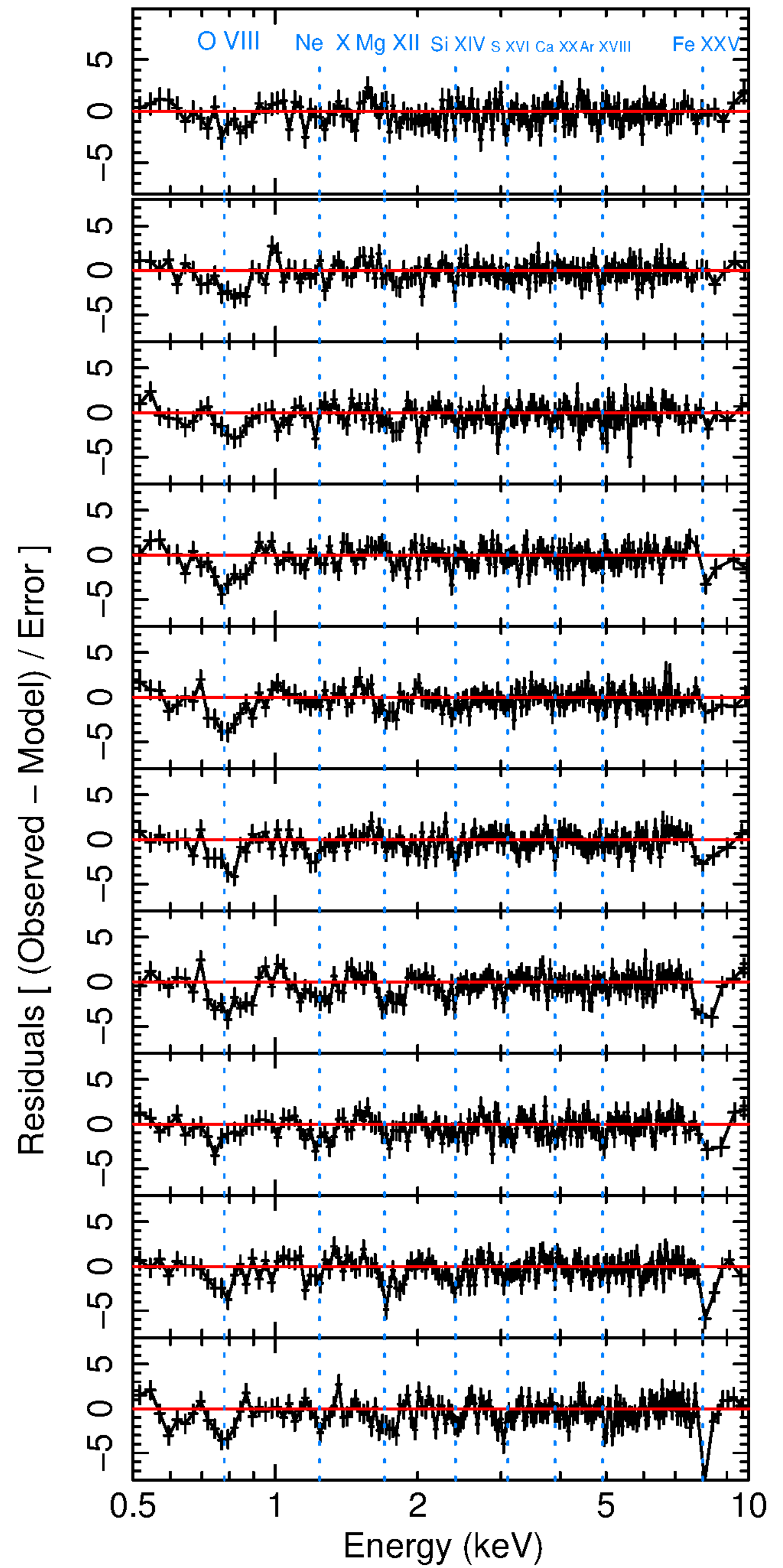}
  \centering
  \vspace{-0.5cm}
   \caption{IRAS 13224 flux-selected EPIC/pn spectral fits 
                (ordered according to their luminosity from bottom to top).
                Energy transitions of relevant ions blueshifted by $0.24c$ are labelled.
                {Note the persistent O\,{\scriptsize VIII} absorption line compared
                to the strong decrease of the iron line at high fluxes.}} 
            \label{Fig:Fig_pn_spectra_residuals}
\end{figure}

This continuum model describes the EPIC flux-resolved spectra reasonably, 
but strong residuals are found in terms of absorption features
around 0.8 keV ($\sim16$\,{\AA}) and 8 keV (see Fig.\,\ref{Fig:Fig_pn_spectra_residuals}), 
along with several weaker absorption features,
which are stronger at low flux regimes and 
are mainly due to absorption from the UFO as shown in P17a,b. 
The energies of the residuals agree with the blueshifted ($0.24c$) K-shell transitions
of the most abundant ions in the X-ray energy band, such as O\,{\scriptsize VIII}, 
Ne\,{\scriptsize X}, Mg\,{\scriptsize XII}, and Fe\,{\scriptsize XXV}. 

{Interestingly, the O\,{\scriptsize VIII} absorption line is still present, though fainter,
at high fluxes, whilst the iron line seems to vanish (see also Fig. \ref{Fig:Fig_pn_spectra}).}
                
Once we have obtained a reasonable description of the source continuum for each of the 10 flux 
regimes, we search for variability in the absorption lines produced by the UFO in the 
soft X-ray band with the high-resolution RGS instruments. This is particularly useful 
since the EPIC-pn lacks the spectral resolution necessary to resolve absorption features below 2 keV.

  \vspace{-0.5cm}

\begin{figure}
  \includegraphics[width=0.975\columnwidth]{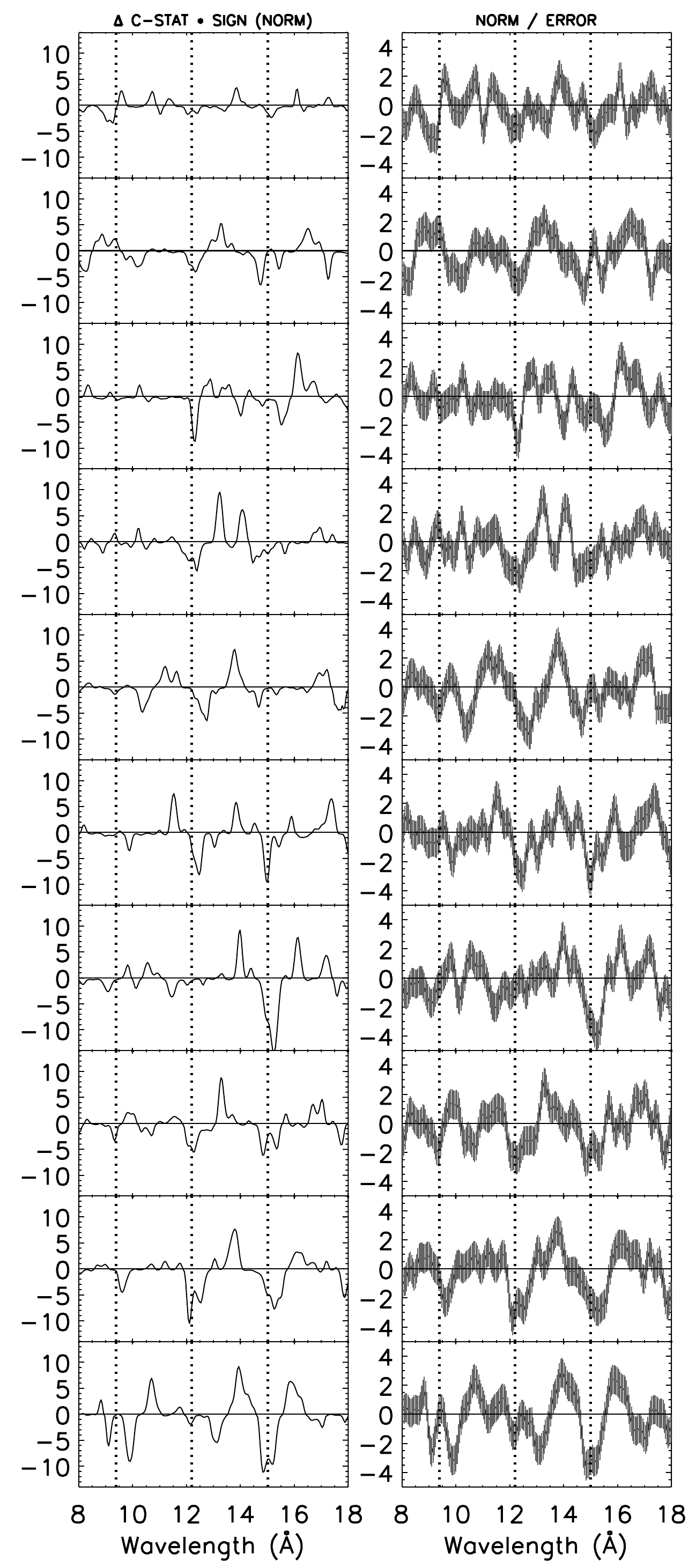}
  \vspace{-0.2cm}
   \caption{RGS-only line search for the 10 flux regimes from low (bottom)
            to high fluxes (top). The left panel shows the $\Delta$\,C-stat multiplied 
            by the sign of the line in order to distinguish both the significance 
            and the kind (emission/absorption) of each feature.
            The right panel shows the ratio between the normalisation of the Gaussian
            and {its error} zoomed on the region with high effective area.
            The bands coloured in grey are the $1\sigma$ error bars. 
            Here the wavelengths are corrected for the redshift (0.0658).
            The vertical dotted lines indicate the rest-frame transitions of 
            the strongest features detected in the 2\,Ms stacked spectrum (P17a).} 
            \label{Fig:Fig_line_search}
\end{figure}

\subsection{The high-resolution RGS spectrum}
\label{sec:rgs_spectra}

At this stage we use the full resolving power of the RGS detector and therefore
do not highly bin the spectra. We adopt bin size equal to 1/3 of the spectral resolution 
for the optimal binning (see Kaastra and Bleeker 2016) and use C-statistics.
We focus on the $6-35$\,{\AA} first order RGS spectra. Below 6\,{\AA} the effective area 
is not well calibrated and above 35\,{\AA} the background is high.

We apply the EPIC continuum model of each flux-resolved spectrum to the corresponding
RGS spectrum with the overall normalisation allowed to vary to account for any uncertainty
in the cross-calibration between the two instruments (in some cases up to 10 per cent).
As for EPIC, the continuum spectral model describes the RGS spectrum well,
although RGS does not have the same broadband sensitivity.
Therefore, the continuum model does not have a {significant} effect on the 
search for narrow lines in the RGS spectra.

Following the approach used in P17a, we search for spectral features 
on top of the continuum by fitting a Gaussian spanning the $6-35$\,{\AA} 
wavelength range with increments of 0.05\,{\AA} and calculate the $\Delta$\,C-statistics.
We test a few different linewidths obtaining consistent results: 
 FWHM $=$ 500, 1000, 5000, and 10,000 km s$^{-1}$. 
We show the results obtained with 5000 km s$^{-1}$
in Fig.\,\ref{Fig:Fig_line_search} because this value is similar to the velocity
width measured in P17a {(FWHM $=2.35 \sigma_v = 2.35 \times 2000 \sim 5000$ km s$^{-1}$)}. 
The wavelengths are corrected for redshift ($z=0.0658$).
The source flux increases from bottom to top. The left panels show the changes in
the $\Delta$\,C-statistics, multiplied by the sign of the normalisation of the Gaussian in order
to distinguish between emission and absorption lines; 
the right panels show the ratio between the normalisation of the Gaussian
and {its error}. 
More detail on use of this line-scan and the comparison between intrinsic and
Galactic absorption lines can be found in Pinto et al. (2016).

As previously seen in P17a for three flux regimes, the 10 flux-resolved spectra show 
significant residuals that weaken at high fluxes.
There is weak evidence of emission-like features,
see Fig.\,\ref{Fig:Fig_line_search}, but they vary both in strength 
and position between the observations for which is difficult to address the significance
and the origin. If they are real they might be related to some reflection
occurring in the disk much further away (around 1000\,$R_S$, see e.g. Blustin and Fabian 2009),
{but their variability would rather suggest that they are produced by reflection 
of coronal photons from the inner accretion disc (Garcia et al. 2016).}
We do not expect these lines to have a major effect onto the measurements 
of the absorption lines and their variability.
The main absorption feature at 15\,{\AA} (or 16\,{\AA}, $\sim0.8$\,keV,  
in the observed frame) detected 
in the low-flux RGS spectra resolves the residuals found in the EPIC spectra 
(see Fig.\,\ref{Fig:Fig_pn_spectra_residuals}). 
         
         
\subsection{EPIC-RGS simultaneous fit: wind model}
\label{sec:simple_wind_model}

As noted in P17a, the absorption features detected in the RGS and EPIC spectra
of IRAS 13224 are not compatible with a rest-frame absorber or a low-velocity
warm absorber commonly found in Seyfert galaxies (see e.g. Krongold et al. 2007 and 
references therein) nor with any absorption from the foreground interstellar medium
(e.g. Pinto et al. 2013 and references therein). 
They are instead all consistent with an outflowing velocity of about 
$0.24c$ as shown by the blue-dotted lines in Fig.\,\ref{Fig:Fig_pn_spectra_residuals},
which agree with an UFO nature.
 
An accurate modelling of a photoionised absorber requires both broadband coverage
and high-spectral resolution. 
We therefore simultaneously fit the 10 flux-selected EPIC-pn and RGS spectra and, 
for consistency, we group the EPIC and RGS spectra to a signal-to-noise ratio
of at least 6 and use $\chi^2$ statistics. We also make sure that the bin size is always of 
1/3 of the spectral resolution or above. 
This spectral binning mainly affects the wavelength range below 10\,{\AA}
and above 27\,{\AA} in the low-flux RGS spectra and does not decrease our resolving power
because the UFO soft features - detectable with RGS - are between 10 and 16\,{\AA}.

As previously done in P17a, we model the absorption features with the {\scriptsize XABS}
model in {\scriptsize{SPEX}}. This model calculates the transmission of a slab of material, where all ionic 
column densities are linked through a photoionisation model. 
It is a flexible code that can reproduce photoionised plasma in different ionisation fields
such as in the interstellar medium and X-ray binaries (e.g. Pinto et al. 2012a,b). 
The relevant parameter is the ionisation parameter 
\vspace{-0.1cm}         
\begin{equation} \label{Eq:Eq_ionpar}
\xi = \frac{L_{\rm ion}}{n_{\rm H} \, R^2}
\end{equation}
with $L_{ion}$ the ionising luminosity, $n_{\rm H}$ the hydrogen number density and 
$R$ the distance from the ionising source (see, e.g., Steenbrugge et al. 2003). 
We apply the {\scriptsize XABS} as a multiplicative component onto all emission components: 

\noindent {\scriptsize \;  HOT * RED * XABS * (BB + POW + LAOR * (GAUSS + GAUSS)) },

\noindent which physically can be described as:

\noindent {\scriptsize \;  ISM * Redshift * Wind * (Continuum + Rel-broad * (Fe L + Fe K)) },

\noindent where the relativistically-broadened Fe L (1\,keV) / K (7\,keV) lines and the blackbody +
power-law continuum are at first absorbed by the photoionised wind, then corrected for redshift,
and finally absorbed by the mainly neutral interstellar medium in the Galaxy.

\begin{figure}
  \includegraphics[width=0.95\columnwidth]{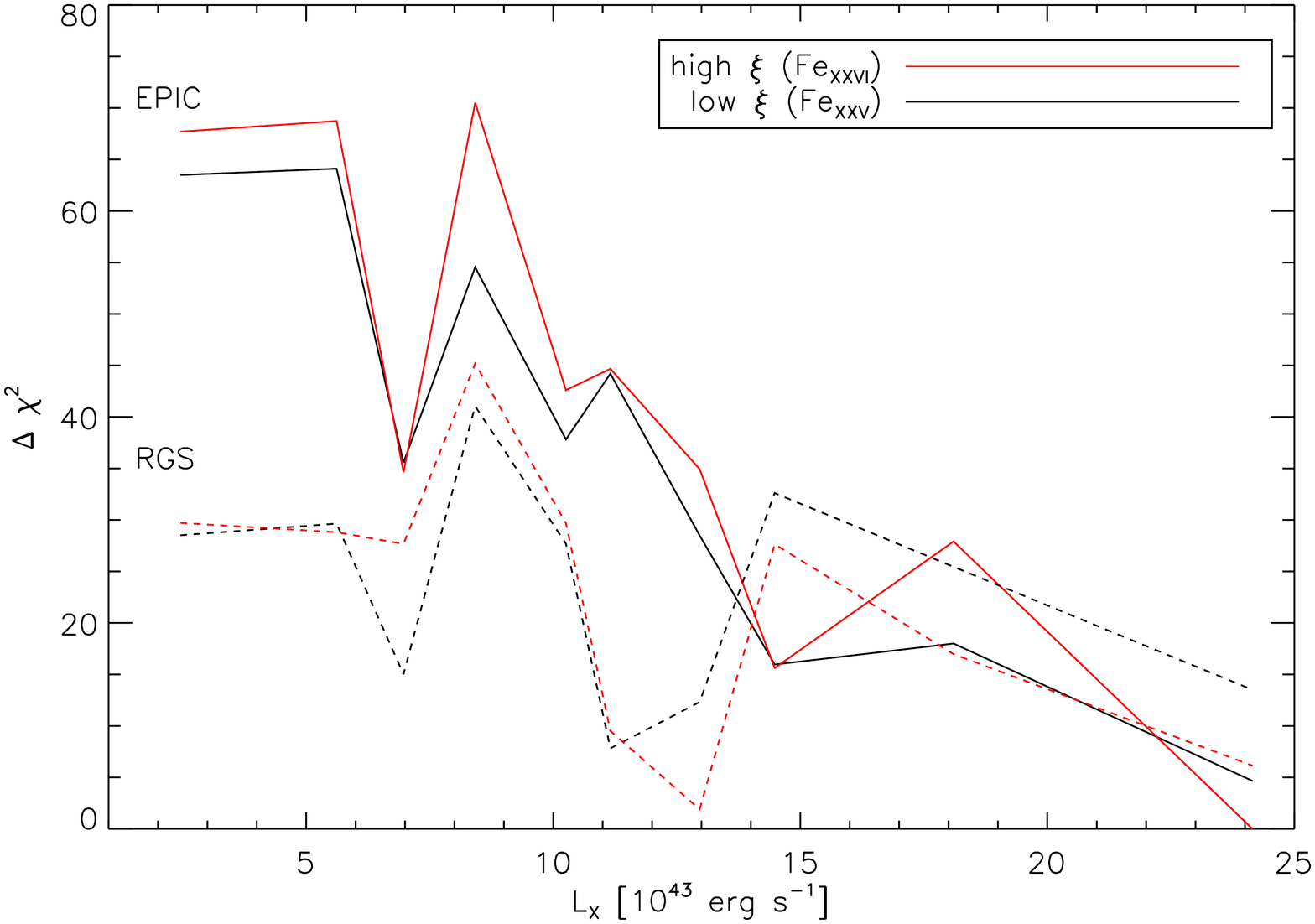}
\vspace{-0.2cm}         
   \caption{Statistical improvements of the absorber obtained on the flux-resolved EPIC
                 and RGS spectra. Both the low $\xi$ and high $\xi$ solutions are shown,
                 where the former one is detailed in Table\,\ref{table:UFO_stat}. } 
            \label{Fig:Fig_xabs_results}
\end{figure}

The free parameters in the {\scriptsize XABS} model are the column density, $N_{\rm H}$,
and the ionisation parameter, $\xi$. The linewidth, $\sigma_v$ and the line-of-sight velocity
$v_{\rm LOS}$ are tied to the values measured in P17a ($2000$ {\kms}, $0.236c$ or $0.21c$), 
with the more sensitive 2\,Ms archival+new data stacked RGS-EPIC spectra. 
The faster ($0.236c$) lower-ionisation solution - 
where the Fe {\scriptsize XXV} significantly contributes to the absorption at 8\,keV - 
is slightly preferred by the 2\,Ms stacked spectrum (see P17a)
and by the PCA analysis (P17b),
but the alternative slower ($0.21c$) higher-ionisation model also provides reasonable fits.
We therefore perform the following analysis for both solutions   
for completeness. 
{At a first stage, we fixed the LOS velocities to the values measured in P17a 
with the time-averaged spectrum which has much higher S/N than each flux-resolved spectrum.}
In Table \ref{table:UFO_stat} we report the $\chi^2$ values 
for each EPIC and RGS flux-resolved spectrum from the simultaneous fits (fast model)
compared to the results obtained with the continuum model.
The statistical improvements by the photoionised absorption models
are plotted in Fig.\,\ref{Fig:Fig_xabs_results} individually for 
EPIC and RGS (both fast and slow models).
{Spectra with best-fit models and residuals are shown in Fig. \ref{Fig:Fig_pn_rgs_spectra} 
and \ref{Fig:Fig_pn_rgs_residuals}. The fits of the $0.236c$ and $0.21c$ solutions are remarkably similar.
The parameters of the absorbers are reported in Table\,\ref{table:table_absorbers}.
We have confirmed the variability of the absorber with additional
fits where the absorption model was fixed to that of the lowest flux spectrum 
and the continuum was free (see Fig. \ref{Fig:Fig_pn_rgs_spectra_model1} 
and \ref{Fig:Fig_pn_rgs_residuals_model1}). 
Deviations are found corresponding to the main transitions 
(blue dotted lines). 
We report the detail of the adopted spectral energy distribution and the ionising flux
in Appendix \ref{sec:appendix} and Figs. \ref{Fig:Fig_SED} and \ref{Fig:Fig_stability_curves}.}
 
Most absorption features in each EPIC and RGS flux-resolved spectrum can be well
described with one photoionised {\scriptsize XABS} absorber with a high ionisation parameter 
$(\log \xi \sim 4-5)$ and a relativistic outflow velocity of $0.236c$, very similar to the 
results obtained in P17a,b. 
Interestingly, we do not find significant deviations in the column density of the absorber 
between the 10 different flux-resolved spectral fits.
The values measured with the Fe {\scriptsize XXV} and Fe {\scriptsize XXVI} solutions
are consistent with those reported in P17a, albeit with larger uncertainties due to the fact
that we split the entire observation in 10 flux intervals. 

In Jiang et al. we also use deeper spectra splitting the data in just three flux intervals 
and found tentative evidence of a second photoionised absorber,
which marginally improves the fit of the Fe K blue wing.
This absorber is not detected in our spectra due to their much shorter exposure and 
is not expected to affect our results.

For completeness, we have tested the effect of any warm absorber through
the time-integrated stacked RGS spectrum (see P17a) and
fitted it with the same continuum and UFO
model used in this paper and have then added a low-velocity 
($0-5000$ {\kms}) photoionised absorber 
with log $\xi \sim 1-4$ (an additional {\scriptsize XABS} component).  
The addition of the warm absorber does not significantly improve the fit. 
We have also tried to remove the UFO model and fit the absorption features 
just with the warm absorber, but that is unsatisfactory 
due to the fact that no additional {\oviiviii} absorption features are detected 
as required from a standard warm absorber model.
{The overall improvement of the warm absorber model (for fits without the UFO model)
is $\Delta C/dof=20/3$, compared to the $>150/3$ for the UFO,
and $\Delta C\sim$ few for each flux resolved spectrum.}
We obtain a 90 per cent upper limit on the column density 
$N_{\rm  H}^{\rm WA} \lesssim 2 \times 10^{20}$\,cm$^{-2}$ for log $\xi \sim 2$. 
Similar results are given by a low-velocity emission component.

 \begin{table}
 \caption{Statistical improvements of the absorber}  
 \label{table:UFO_stat}      
 \renewcommand{\arraystretch}{1.1}
  \small\addtolength{\tabcolsep}{1pt}
  
 \scalebox{1}{%
 \begin{tabular}{c c c c c c c c c c}     
 \hline  
                 &  \multicolumn{3}{c}{EPIC-pn}     & \multicolumn{3}{c}{RGS} \\
                 &  \multicolumn{2}{c}{Continuum}  & Abs   & \multicolumn{2}{c}{Continuum}  & Abs  \\
 Level       &  $\chi^2$  & d.o.f.  &  $\Delta \chi^2$  &  $\chi^2$  & d.o.f.  &  $\Delta \chi^2$ \\
 \hline                                                                                   


  1     &  328  &  177    &   64     &   377   &  335   &   28      \\
  2     &  254  &  171    &   64     &   361   &  322   &   29      \\
  3     &  212  &  166    &   36     &   448   &  357   &   15      \\
  4     &  258  &  167    &   54     &   417   &  364   &   41      \\
  5     &  193  &  163    &   38     &   418   &  348   &   27      \\
  6     &  203  &  167    &   44     &   433   &  371   &   7      \\
  7     &  210  &  158    &   29     &   468   &  379   &   12      \\
  8     &  229  &  162    &   16     &   460   &  353   &   33      \\
  9     &  185  &  164    &   18     &   473   &  386   &   25      \\
 10    &  174  &  162    &     5     &   451   &  392   &   13      \\

 \hline                
 \end{tabular}}
 
Notes: each photoionised absorber has free parameters the column density
and the ionisation parameter (Fe\,{\scriptsize XXV} solution). 
The velocity broadening and the $v_{\rm LOS}$ are fixed to P17a value.
We show the $\Delta \chi^2$ trends for the Fe\,{\scriptsize XXV-XXVI} solutions 
in Fig.\,\ref{Fig:Fig_xabs_results}.

\vspace{-0.25cm}         
 
 \end{table}

\vspace{-0.5cm}         
 
\subsubsection{Wind velocity versus X-ray luminosity}

\begin{figure}
  \includegraphics[width=0.95\columnwidth]{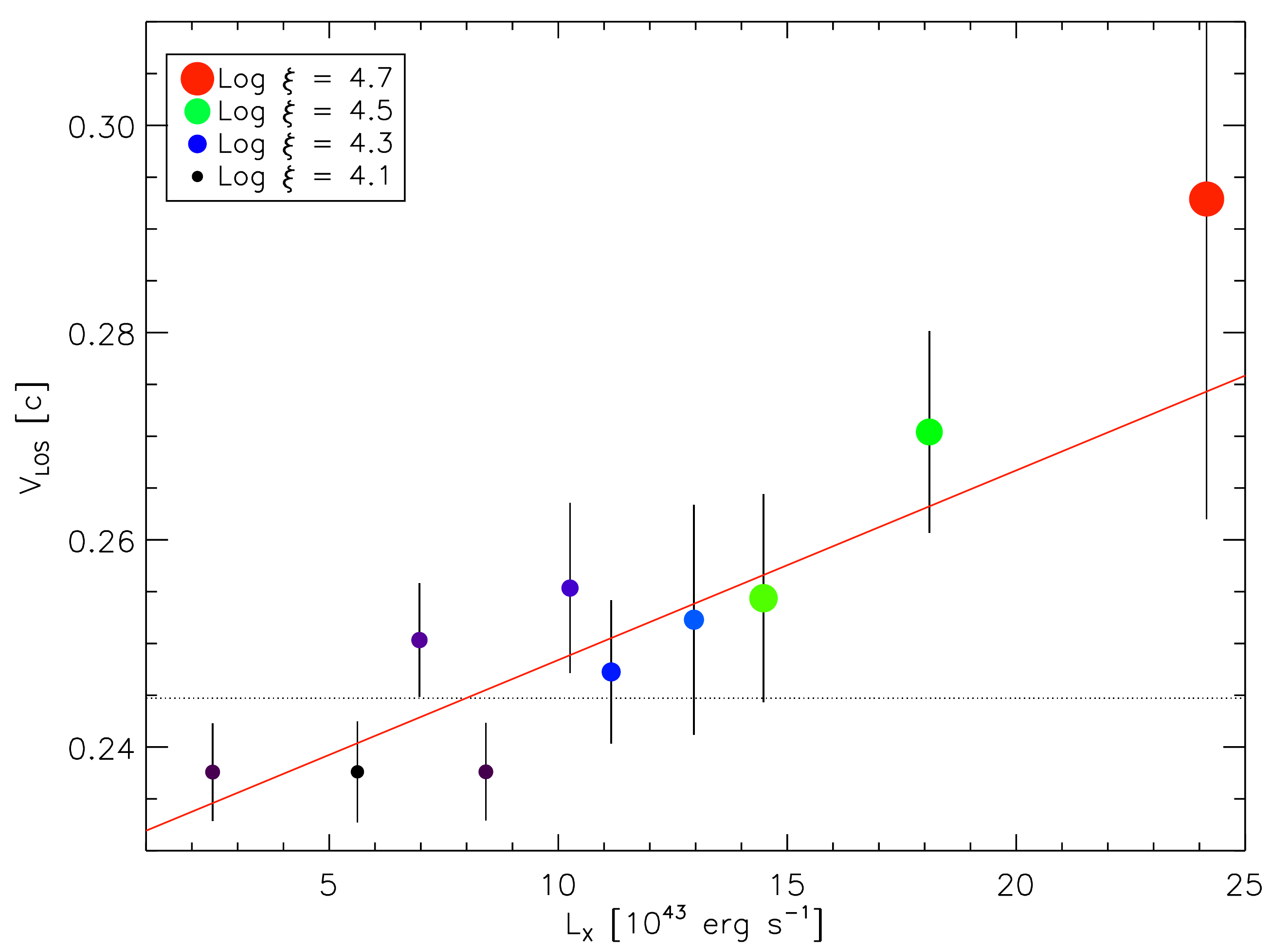}
\vspace{-0.25cm}         
   \caption{Line-of-sight velocity of the wind versus X-ray luminosity. 
                 The point colour and size are coded according to the 
                 value of the ionisation parameter. The red solid line shows a linear fit.
                 The horizontal line is the fit with a constant function.} 
            \label{Fig:Fig_zv_xlum_results}
\end{figure}

Radiation-driven winds are expected to show correlation between the outflow speed
and the X-ray luminosity (see, e.g., Matzeu et al. 2017 and references therein).
We have therefore performed a fit with free velocity among the 10 spectra,
which shows velocities broadly consistent within the error bars 
(see Fig.\,\ref{Fig:Fig_zv_xlum_results}, {for the faster solution}).
A constant function yields $\chi^2_{\nu} = 2.26$ for 9 degrees of freedom
 ($p$-value $=$ 0.016),
slightly improved by a velocity-luminosity linear fit ($\chi^2_{\nu} = 0.90$
for 8 degrees of freedom, $p$-value $=$ 0.52, {and a slope $0.0019\pm0.0005$}). 

The Pearson and Spearman correlation coefficients of the $(v,L_X)$ points
are 0.93 and 0.86, respectively, implying a strong correlation. 
We establish the significance of this correlation using a Monte-Carlo approach. 
We simulate $10^6$ equivalent sets of points, assuming that the velocity of the outflow is constant, 
and that the errors are Gaussian. Of these, 483 have a higher Pearson coefficient, 
3595 have a higher Spearman coefficient, and 418 have both higher. 
These give significances of 99.95\%, 99.64\%, and 99.96\%, respectively.
Our results agree with the trend observed in
the quasar PDS 456 and a scenario of radiation-driven wind (Matzeu et al. 2017).

{There is however some degeneracy between the two solutions 
(high Fe\,{\scriptsize XXV} versus high Fe\,{\scriptsize XXVI}) 
at this stage. 
The velocity separation between the two scenarios are within the error bars of 
($\sim 0.01c$ statistical plus the line broadening $\sim5000$ km s$^{-1}$). 
{The peaks of the $\chi^2$ distributions are separated by a few digits 
for some flux intervals (see e.g. Fig.\,\ref{Fig:Fig_xabs_results}), which weakens the trend
and therefore we consider it just as tentative.}
It is possible that the flux-selection of events at different epochs 
somehow smeared out the results. 
In a forthcoming paper we will try to obtain more information on the 
source variations through time-resolved spectroscopy.}

\subsubsection{Wind ionisation versus X-ray luminosity}

The goal of this work is to understand the behaviour of the wind across different flux states
and search for any luminosity dependence. 
As can be seen in Table \ref{table:UFO_stat} and Fig.\,\ref{Fig:Fig_xabs_results},
the addition of a photoionised absorption model improves most of the fits
whether we adopt the low or high $\xi$ solutions. 
However, the largest difference in $\chi^2$ are to be found at lower flux states,
which confirms that the wind features weaken with increasing flux.
 
\begin{figure*}
  \includegraphics[width=1.7\columnwidth]{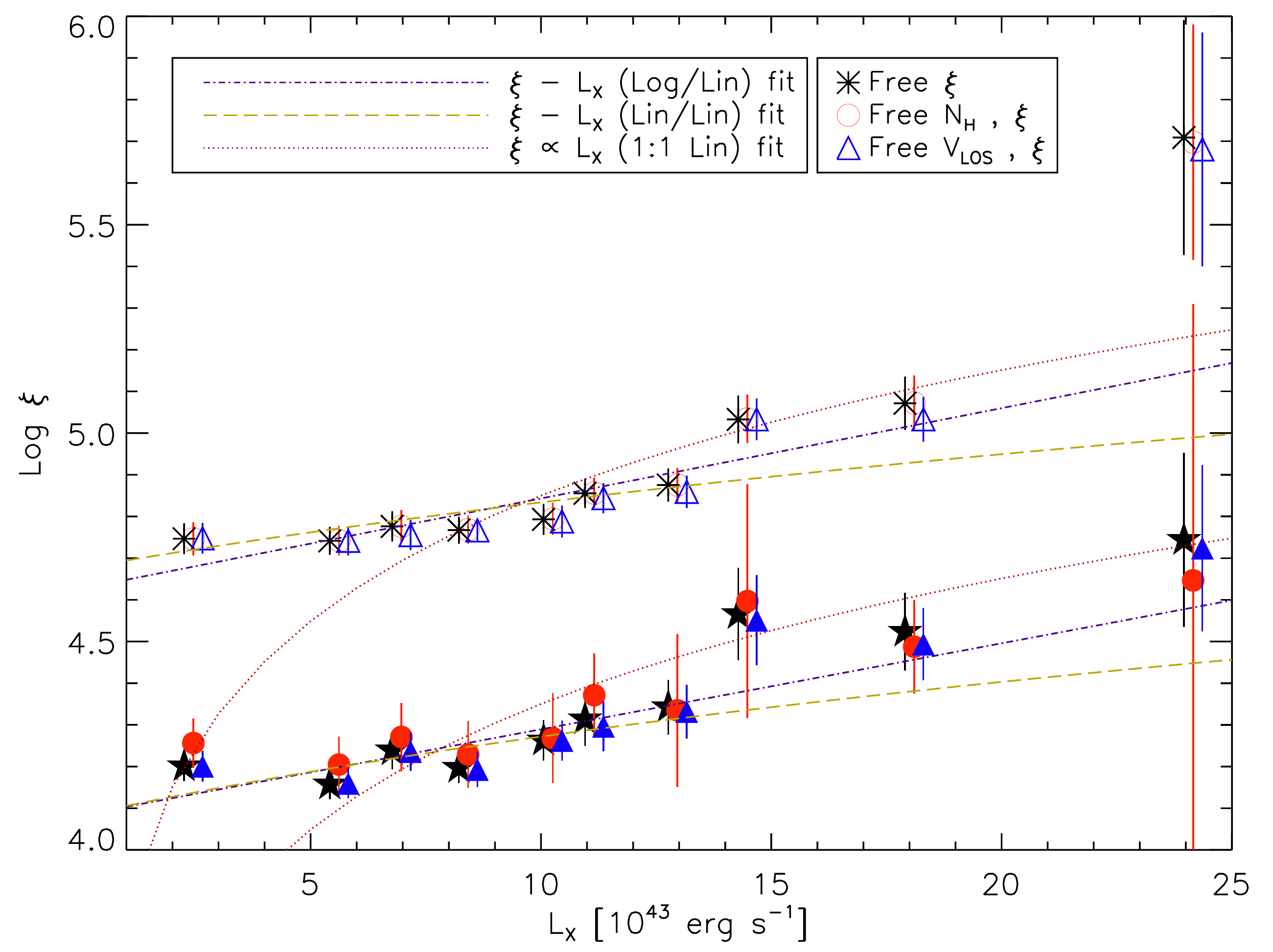}
   \vspace{-0.3cm}            
   \caption{EPIC/pn+RGS results: ionisation parameter versus X-ray luminosity.
                The results of both the high $\xi$ (open points) and low $\xi$ 
                (filled points) models are shown along with
                their liner and log-linear fits (see Eq.\,\ref{Eq:Eq_linefit}).
                Stars, circles, and triangles indicates spectral fits with different free parameters.
                Log\,$\xi-L_X$ and $\xi-L_X$ fits are performed. 
                The points clearly deviate from the 1:1 (dotted) line.} 
            \label{Fig:Fig_xabs_vs_xlum}
   \vspace{-0.3cm}            
\end{figure*}

In Fig.\,\ref{Fig:Fig_xabs_vs_xlum}, we show the ionisation parameter measured
for each RGS-EPIC flux-selected spectrum and the corresponding X-ray luminosity 
estimated between 0.3--10\,keV totalling the contributions of the four emission components.
We prefer to use the luminosities as measured with the continuum fits because they are not
affected by the photoionisation modelling and provide lower limits to the actual source brightness.
Both the Fe {\scriptsize XXV} (filled circles) and Fe {\scriptsize XXVI} (open circles) solutions are shown.
We also show the results where the line-of-sight velocity is free to vary (triangles)
and where the column density is tied to the P17a value (stars).

All the spectral fits show a clear correlation between the ionisation parameter
and the X-ray luminosity. The origin of such trend is discussed in the following section.
Here we mainly notice that a linear fit to the $\xi - L_X$ points provides:

\begin{eqnarray} \label{Eq:Eq_linefit}
\;\;\; \frac{\xi}{10^4} = (4.7\pm0.8) + (0.21\pm0.06) \frac{L_X}{10^{43}} \;\;\; (\rm Fe_{\,XXVI})\\
\;\;\; \frac{\xi}{10^4} = (1.2\pm0.3) + (0.07\pm0.02) \frac{L_X}{10^{43}} \;\;\; (\rm Fe_{\,XXV}) \;
\end{eqnarray}
corresponding to $\chi^2_{\nu} = 2.7$ and 1.4, respectively, for 8 degrees of freedom.
A constant function yields significantly worse fits, $\chi^2_{\nu} = 5.8$ and 4.1, respectively.

The highest-flux pn spectra are just above the pile-up threshold at energies below 1 keV, 
potentially affecting the spectrum up to 2 keV. 
We have repeated the simultaneous RGS and EPIC spectral fits, 
ignoring the EPIC data below 2 keV and covering the 0.4--2 keV band with RGS only. 
This test provides results on the {\scriptsize XABS} parameters consistent with those 
obtained using the full 0.3-10\,keV EPIC spectra, which suggests that pile up has minor
effects on our analysis.

\vspace{-0.5cm}         

\section{Discussion}
\label{sec:discussion}

In P17a we have discovered mildly-relativistically blue-shifted absorption features 
in the \textit{XMM-Newton} and \textit{NuSTAR} spectra of IRAS\,13224, 
which were interpreted as strong evidence for an ultrafast outflow. 
We further found that the features weaken when the flux increases, which 
can be explained as a decrease of the column density or an increase of the ionisation 
parameter. More recently, the PCA analysis of IRAS\,13224 (P17b) has provided 
further support for the latter scenario showing, however, that the response of the 
absorber is not as strong as the luminosity variations. 

\vspace{-0.5cm}         

\subsection{High Eddington accretion?}

Ultrafast winds are expected to be driven by radiation pressure when the compact object
accretes at or above the Eddington limit (see e.g. Takeuchi et al. 2013). 
But at which Eddington rate is IRAS\,13224 actually accreting? 

The exact mass of the supermassive black hole powering IRAS\,13224 is not well known.
For a black hole with a mass of $M_{\rm BH} = 10^7 M_{\odot}$, 
which is a conservative value for our source and NLS1 in general (see e.g. Chiang et al. 2015), 
the Eddington limit $L_{\rm Edd} = 3.2 \times 10^4 (M_{\rm BH} / M_{\odot}) L_{\odot}$
is reached at about $1.3 \times 10^{45}$ erg s$^{-1}$.

We can estimate the bolometric luminosity through the 
hard X-ray (2--10 keV) to bolometric luminosity correction according to
Vasudevan \& Fabian (2007). They reported correction factors between 15 and 70
for NLS\,1 galaxies, which would suggest that IRAS\,13224 bolometric luminosity
spreads between 0.02 and $1.6 \times 10^{45}$ erg s$^{-1}$, with an average
throughout the \textit{XMM-Newton} observations of $\sim 3 \times 10^{44}$ erg s$^{-1}$.
Despite systematic uncertainties this suggests that the source accretes at significant 
fraction of the Eddington limit and possibly surpasses it during the brightest epochs.

In a parallel paper that focuses on the continuum decomposition and variability
of IRAS 13224 (Jiang et al.) we use the data available on the 
NASA/IPAC Extragalactic Database\footnote{https://ned.ipac.caltech.edu} for the source
from optical to hard X-rays to model the spectral energy distribution.
{The optical to hard X-ray SED is fitted with a blackbody plus powerlaw emission model.}
We estimate a range for the bolometric luminosity of $\sim0.3-3.0\,L_{\rm Edd}$
{(with the uncertainty mainly driven by the blackbody temperature adopted)}.
This agrees with Sani et al. (2010) and the average measurements of Buisson et al. (2017)
of about 70 per cent Eddington.
{More recently, Buisson et al. (2018) obtained tighter limits on the luminosity for the 
time-average SED ($\sim0.3-1\,L_{\rm Edd} \times \frac{10^7\,M_{\odot}}{M_{\rm BH}}$) 
using an additional blackbody for the UV emission from thin disc.}

We notice that all the previous calculations make use of the observed luminosity.
This will be significantly lower than the intrinsic luminosity which has to account
for any X-ray absorption and scattering from the ultrafast wind. Our approach is conservative 
and further supports the chance that IRAS\,13224 accretes beyond the Eddington limit.

We have found a correlation between the velocity
of the wind and the X-ray luminosity (see Fig.\,\ref{Fig:Fig_zv_xlum_results}).
Our results agree with the trend observed in
the quasar PDS 456 and with the scenario of a radiation-driven wind 
at high accretion rates (see, e.g., Matzeu et al. 2017 and references therein).

\vspace{-0.5cm}         

\subsection{Scenario 1 -- wind disappearance at high flux}
\label{sec:discussion_ionvslum}

One of the main results from the RGS line search is that the absorption lines weaken 
and seem to disappear at higher luminosities (see Fig.\,\ref{Fig:Fig_line_search}).
In Fig.\,\ref{Fig:Fig_xabs_vs_xlum} we compare the ionisation parameter measured 
for the photoionised absorber with the corresponding luminosity of the 10 flux-resolved spectra.
Whether we adopt the Fe {\scriptsize XXV} or the Fe {\scriptsize XXVI} solutions 
we obtain a clear correlation between the ionisation parameter
and the X-ray luminosity. 
This trend may suggest that when the source brightens, possibly due to higher accretion, 
the enhanced radiation field increases the ionisation state of the wind, which would result
in a decrease of the EWs of the X-ray lines at the same column density of the wind.
It is however suspicious that the $\xi-L_X$ relationship is not a straight line as expected
from Eq.\,\ref{Eq:Eq_ionpar} for constant wind geometry. 
In particular the slope is less than unity,  $\xi \propto 0.05-0.3 \; L_X$ (see Eq.\,\ref{Eq:Eq_linefit}-3), 
which could suggest that the wind is either not
responding on time or that something else is preventing the wind to adjust its ionisation
state to the luminosity variations. 
It is possible that only a certain fraction of the luminosity is actually ``seen" by the wind
and that a significant fraction is produced in an outer, screened, region.
We notice that there are some systematic uncertainties 
in this comparison due to the fact that we are using the observed luminosities rather than intrinsic, 
unabsorbed, luminosities which are not known a-priori.

{Naively, the Fe {\scriptsize XXVI} solution would better follow the disappearance of the lines 
at higher ionisation states. However, the same column density for Fe {\scriptsize XXVI} would provide absorption lines weaker than for Fe {\scriptsize XXV} due to the cross sections. The fact that the 
lines do not ``disappear" until the highest flux states are reached and that the O {\scriptsize VIII} is always present would suggest the opposite.}
 
{We have chosen to compare the ionisation parameter, $\xi$, with the full-band X-ray luminosity as 
there are absorption features over the whole 0.3$-$10 keV band. However, it is interesting to compare
the $\xi$ with the luminosity of the harder band, such as 5$-$10 keV, which contributes most of the Fe K absorption.
The variations on the luminosity are smaller, approximately twice less than the total band
and, interestingly, the corresponding $\xi-L_X$ trend 
{can be fitted with a straight line of slope $2-4$}.
This would suggest that some of the soft X-ray emission may not be seen from the part of the wind which
affects our line of sight, particularly since the O {\scriptsize VIII} is always present.}

If the wind geometry and location do not change, we can use Eq.\,\ref{Eq:Eq_ionpar}
to place constraints on the density of the gas, which is useful to calculate the recombination 
time scales of the wind. If we assume that the velocity of the wind ($\sim0.2c$) is equal to the 
escape velocity, then we obtain a launching radius of about 25 Schwarzschild radii, $R_S$.
Using Eq.\,\ref{Eq:Eq_ionpar} and the luminosity range of IRAS\,13224 as shown in 
Fig.\,\ref{Fig:Fig_xabs_vs_xlum}, we estimate a number density 
$n_{\rm H} \gtrsim 10^{10}$\,cm\,$^{-3}$. 
Using the ionisation parameters estimated for the {\scriptsize{XABS}} model and
the {\scriptsize{SPEX REC$\_$TIME}} tool, we measure the product 
$n_{\rm H} \cdot t_{\rm rec}$ between the density and the recombination time 
for the most abundant ions detected in the wind such as O\,{\scriptsize{VIII}}, 
Mg\,{\scriptsize{XII}}, Si\,{\scriptsize{XIV}}, and Fe\,{\scriptsize{XXV}}. 
The ratio between the $n_{\rm H} \cdot t_{\rm rec}$ provided by {\scriptsize{REC$\_$TIME}} 
and the density, $n_{\rm H}$, yields an upper limit on the recombination time
of a few seconds. This means that the wind should be able to respond almost 
instantaneously to the luminosity variations.

Interestingly, Gallo et al. (2004) reported doubling time scales between 300--800s during which 
IRAS 13224 flux change significantly ({and likely even shorter 
in the new data, e.g. Jiang et al.}). This means that at $25R_S$, which corresponds
to a light travel time of 2500s, the variations in the ionising state might start to blur together
and weaken the $\xi-L_X$ correlation. If the absorber is closer to the ionisation source
the correlation should strengthen, whilst higher distances would weaken it further.
 
\vspace{-0.5cm}         

\subsection{Scenario 2 -- clumpy wind or variable absorption from the disc photosphere?}
\label{sec:discussion_nhvslum}

Alternatively, it is possible that the absorption lines weaken at high luminosities due to
a smaller column density of the intervening absorber. This could happen either if the 
wind becomes more clumpy or if line of sight of the absorber changes with respect 
to that towards the X-ray source.

\vspace{-0.5cm}         

\subsubsection{The column density changes in a clumpy wind}

If a variation in the column density, $N_{\rm H}$, of the wind could explain both the change in 
luminosity and in line strength, then we should be able to fit all the flux-resolved EPIC+RGS spectra 
assuming the continuum model from the brightest flux state and fit the $N_{\rm H}$ of the 
photoionised absorber, of course expecting to find higher $N_{\rm H}$ for fainter spectra.
We have tested this scenario freezing continuum models 
but obtained bad fits ($\Delta \chi^2 \gtrsim 100$ each) due to 
the fact that the spectral shape of IRAS\,13224 weakly changes from the brightest
to the lowest flux (see Figs.\,\ref{Fig:Fig_pn_spectra} and \ref{Fig:Fig_rgs_spectra}).
We should expect spectra significantly harder at higher $N_{\rm H}$.

\begin{figure}
  \includegraphics[width=0.975\columnwidth]{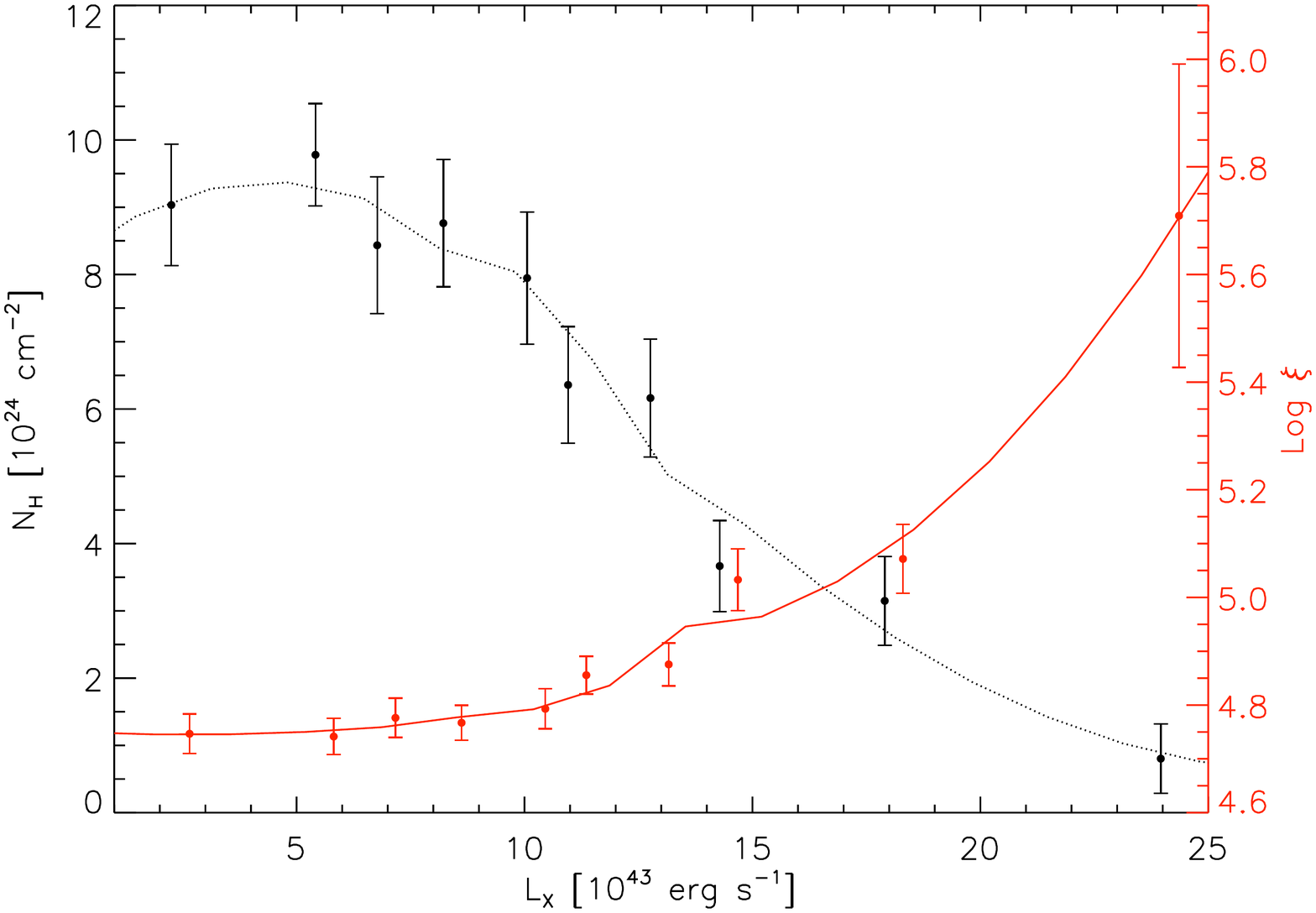}
\vspace{-0.3cm}         
   \caption{EPIC/pn+RGS results: column density versus luminosity 
               (black, fixed ionisation parameter)
                compared to ionisation parameter versus luminosity (red).
                For high $\xi$, see also Sec.\,\ref{sec:discussion_nhvslum}.
                The lines show a quadratic interpolation to both the relations.
                The $N_{\rm  H}$ fits are slightly worse than the $\xi$ fits by $\Delta \chi^2 \sim 10-50$.} 
            \label{Fig:Fig_nh_vs_xlum}
\vspace{-0.3cm}         
\end{figure}

If the ionisation parameter of the wind does not change then both its column density and
the source luminosity have to change in order to have acceptable fits. 
The column density has to decrease precisely with the luminosity.
We have tested this case by fitting again the 10 flux-resolved EPIC+RGS spectra
by freezing the (log) ionisation parameter to the average value of 4.84 obtained for the 
Fe\,{\scriptsize{XXVI}} solution (see e.g. Fig.\,\ref{Fig:Fig_xabs_vs_xlum}). 
The continuum parameters and the column density of the absorber were free to vary.
These fits are slightly worse than our standard free $\xi$ fits with $\Delta \chi^2 \sim 10-50$ each
throughout the 10 spectra. 
The resulting $N_{\rm H} - L_X$ trend for the 10 flux-resolved spectra is shown in 
Fig.\,\ref{Fig:Fig_nh_vs_xlum} along with the previous results obtained with free $\xi$ and $N_{\rm H}$.
As expected, as the source becomes brighter a decrease in the column density is required 
to describe the weakening of the absorption lines, 
providing an alternative scenario to the increase of the ionisation parameter.

\vspace{-0.5cm}         

\subsubsection{Variable absorption from the disc photosphere}

\begin{figure*}
  \includegraphics[width=1.9\columnwidth]{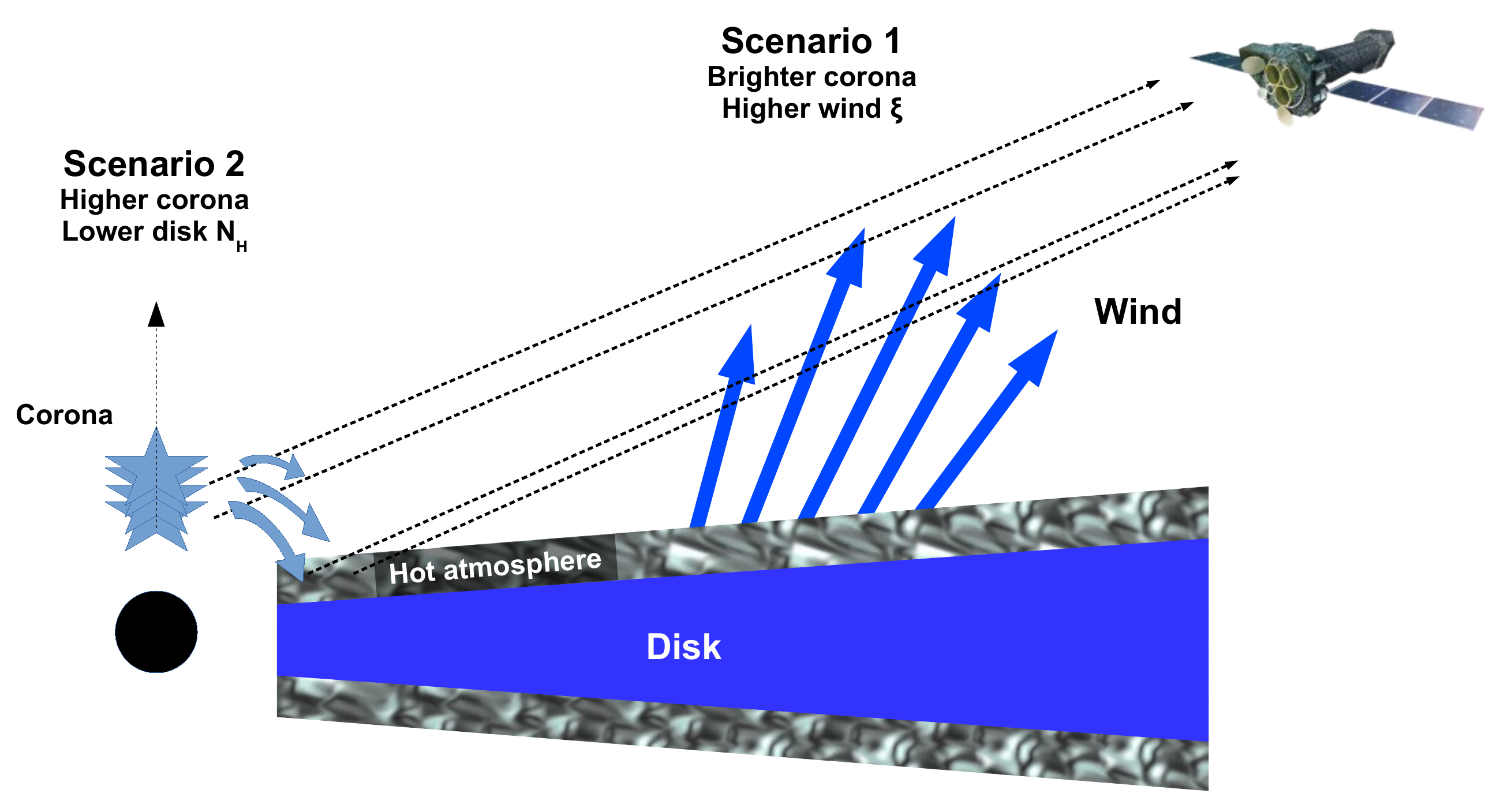}
   \caption{Possible scenarios of outflow response to the flux changes.
                Scenario 1: the X-ray source brightens, it over-ionises the wind whose absorption lines weaken.
                Scenario 2: the average height of the corona increases and the corona brightens due to less 
                light-bending, the disc reflection
                decreases and the corresponding absorption from the photosphere weakens
                compared to the relative source brightening.} 
            \label{Fig:Fig_geometry}
\vspace{-0.25cm}         
\end{figure*}

Gallo and Fabian (2011, 2013) have shown that UFO-like absorption features can be produced 
if an optically thin photosphere on top of the disc absorbs the X-rays from the inner regions
where most of the Fe-K reprocessing (or reflection) occurs.
According to this scenario the absorption applies onto the reflection (secondary) 
rather than the power-law (primary) component,
with blueshifts caused by velocities expected in the inner accretion disc.
In this case, an increase in the height of the corona would decrease light bending,
resulting in a smaller reflection fraction and therefore weaker absorption lines
(see Fig.\,\ref{Fig:Fig_geometry} and Table\,\ref{table:table_continuum}). 

{The actual test for this interpretation requires however an accurate calculation 
of the absorption through the disc atmosphere and will be done elsewhere}. 
For the remainder of the discussion we will assume that the absorption lines
are actually produced by a fast wind rather than by the disc photosphere.


\subsection{Implications for AGN feedback}
\label{sec:discussion_windpower}

Ultrafast outflows could provide an important contribution to AGN feedback and 
significantly affect the evolution of the host galaxy. Their opening angle 
and ejected mass are likely larger than in jets,
allowing them to couple more efficiently to the ISM.
According to Tombesi et al. (2010) UFOs should have a covering fraction of $\sim0.4-0.6$
{as shown in detail for PDS 456 by Nardini et al. (2015)}.
However, their signatures seem to vary in multi-epoch deep observations
with a possible transient nature (e.g. Dadina et al. 2005, Vaughan and Uttley 2008,
Dauser et al. 2012, Zoghbi et al. 2015). 
The exact nature of their variability is not fully understood. 
Some models include variable column density or covering fraction 
(e.g. Reeves et al. 2014, Hagino et al. 2016).

In this work we have shown that the UFO variability in IRAS\,13224 -- 
whose broadband spectrum does not change dramatically within almost 2 orders
of magnitude of X-ray luminosities -- can be attributed to variations in the ionisation parameter
due to the different ionising field.
This is consistent with our recent work on principal component analysis (P17b).
The high ionisation of the outflowing gas explains why it is difficult to detect the UFO
in the bright states of IRAS\,13224 and could potentially be one of the main reasons
for which it is commonly difficult to detect UFOs in AGN.
This suggests that we may currently underestimate their effects onto the
surrounding medium through the process of feedback (see also Gaspari and Sadowski 2017),
and that flux-resolved X-ray spectroscopy can be used to better detect 
and constrain the wind.

In a forthcoming paper we will extend this analysis to the archival data of the brightest
AGN, with particular focus on the NLS1s which provide the statistics
and the different flux states necessary to search for UFOs and study their variability.
We expect future missions such as the X-ray Astronomy Recovery Mission (XARM)
and ATHENA (Nandra et al. 2013) to revolutionise the way we study UFOs thanks to their 
microcalorimeter detectors, which possess high spectral resolution and effective area
throughout the X-ray band.

\vspace{-0.75cm}         

\section{Conclusions}
\label{sec:conclusion}

In this work we have studied the ultrafast outflow recently discovered in 
the narrow-line Seyfert 1 IRAS 13224$-$3809, 
the most rapidly variable AGN in the X-ray energy band.
We have found that the ionisation parameter is correlated with the luminosity
and, {tentatively}, the outflow velocity. 
In the brightest states the gas is almost completely ionised
by the powerful radiation field and the UFO is hardly detected. 
This strengthens our recent results obtained with alternative techniques such as 
principal component analysis.
The high ionisation of the outflowing gas may explain why it is commonly difficult 
to detect UFOs in AGN, suggesting that we may underestimate 
their contribution to AGN feedback. 

\vspace{-0.75cm}         

\section*{Acknowledgments}

This work is based on observations obtained with \textit{XMM-Newton}, an
ESA science mission funded by ESA Member States and USA (NASA).
We also acknowledge support from ERC Advanced Grant Feedback 340442.
WNA acknowledges support from the European Union Seventh Framework 
Programme (FP7/2013-2017) under grant agreement n.312789, Strong Gravity.
DB acknowledges an STFC studentship.
We also acknowledge support European Union's Horizon 2020 Programme 
under the AHEAD project (grant agreement n. 654215).
{We are grateful to Missagh Mehdipour for useful discussion on ionisation balance 
and to the referee, Emanuele Nardini, for his useful comments that improved the
clarity and quality of the paper.}

\vspace{-0.75cm}         



\vspace{-0.5cm}         

\appendix

\section{Additional plots}
\label{sec:appendix}

Here we place some plots and tables that are excluded from the main body of the paper
 to facilitate the reading. 

In Table\,\ref{table:table_continuum} we report our continuum spectral fits to let the reader 
reproduce our results. The broadband EPIC-pn spectra are fitted simultaneously to better
constrain the parameters of the relativistic {\scriptsize LAOR} model which is coupled between
the 10 spectral models. 
Each component is redshifted ($z=0.0658$) and absorbed by interstellar gas 
($N_{\rm H}=6.75\times10^{20}$\,cm$^{-2}$). The two iron lines are also broadened
by a relativistic {\scriptsize LAOR} model with inner radius $ 1.35 \pm 0.01 (GM/c^2)$,
outer radius fixed to 400 $(GM/c^2)$, emissivity slope (or index) $9.28 \pm 0.04$,
and inclination $67.8\pm0.3$ degrees.
We show the residuals to this fits in Fig.\,\ref{Fig:Fig_pn_spectra_residuals}.
We chose to fix the outer radius and to couple the Fe K line energy because the spectra
are not sensitive enough to constrain them well.
The parameters of the {\scriptsize LAOR} component have been fitted simultaneously
to the 10 flux-resolved spectra.

\begin{figure}
  \includegraphics[width=1.02\columnwidth,angle=0]{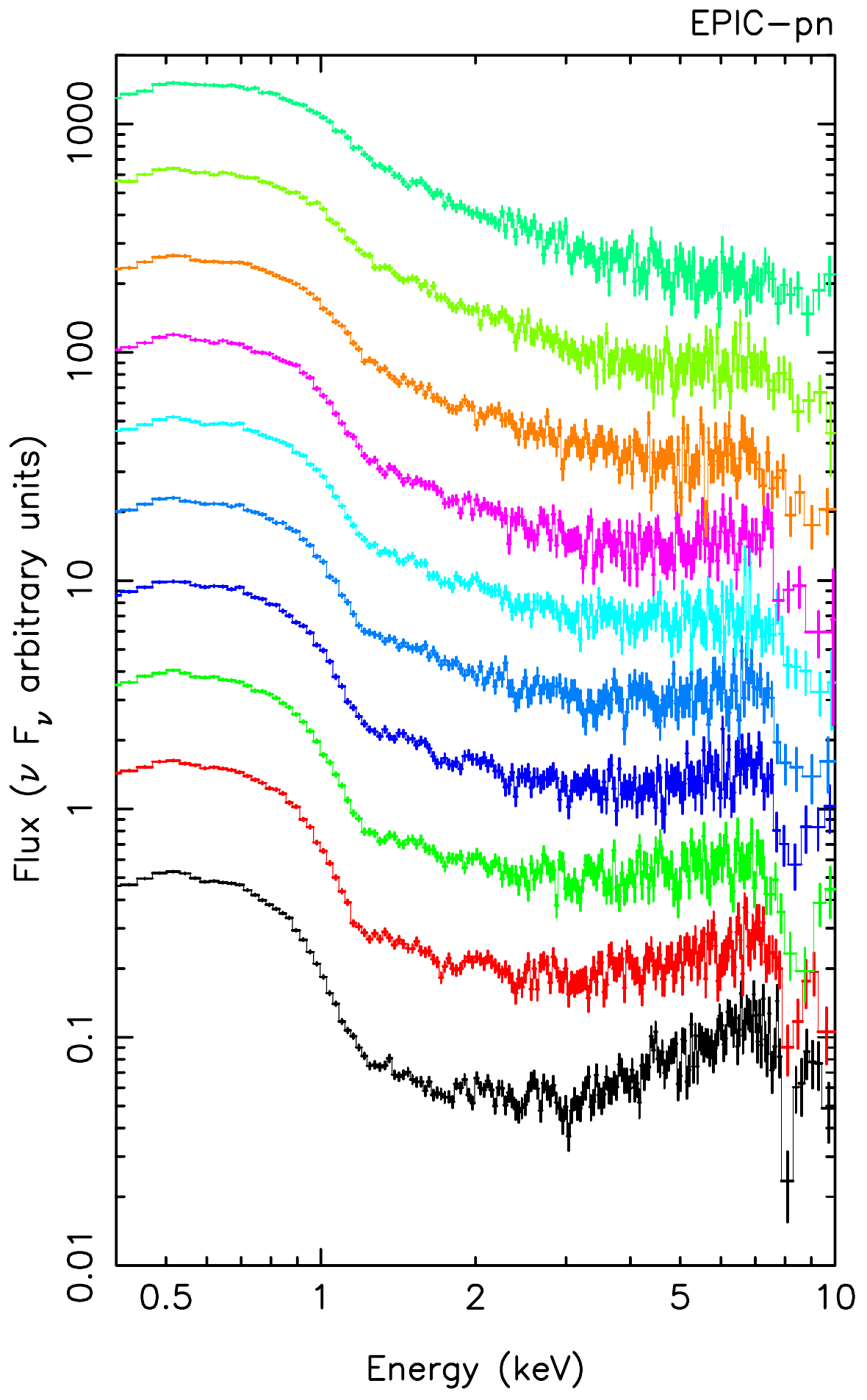}
  \centering
   \caption{IRAS 13224 flux-selected EPIC/pn spectra 
                (ordered according to increasing luminosity from bottom to top).
                The spectra have been shifted along the Y-axis for plotting purposes.} 
            \label{Fig:Fig_pn_spectra}
     \vspace{-0.5cm}       
\end{figure}

\begin{figure}
  \includegraphics[width=1.0\columnwidth,angle=0]{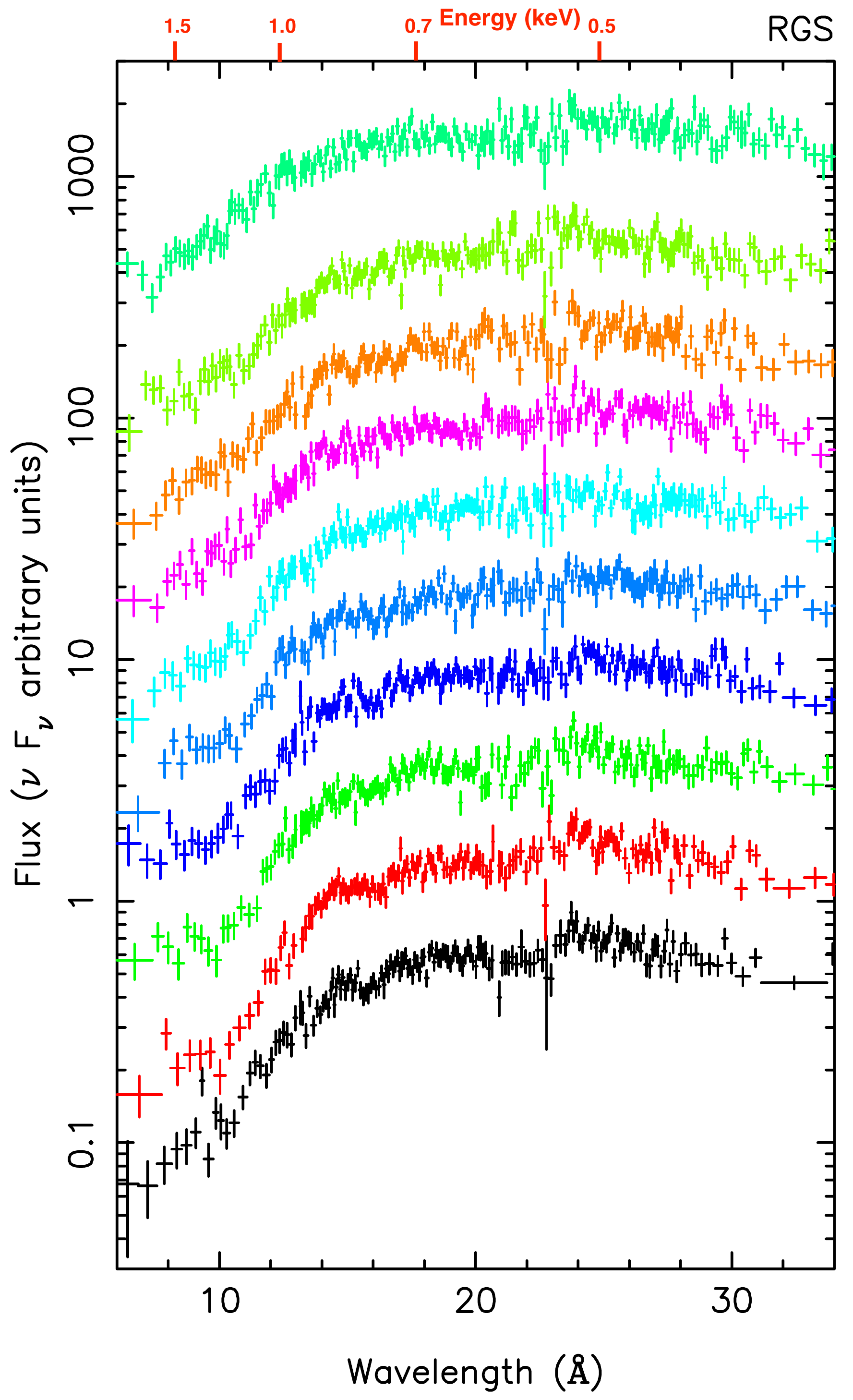}
  \centering
   \caption{IRAS 13224 flux-selected RGS spectra
                (ordered according to increasing luminosity from bottom to top).
                The spectra have been shifted along the Y-axis for plotting purposes.} 
            \label{Fig:Fig_rgs_spectra}
     \vspace{-0.5cm}       
\end{figure}

Figs. \ref{Fig:Fig_pn_spectra} and \ref{Fig:Fig_rgs_spectra} show
the EPIC-pn and the RGS spectra extracted for the 10 flux-resolved intervals ($\nu F_{\nu}$ vs Energy).
They have been multiplied by constant values for display purposes.

We underline that these models are phenomenological and mainly useful to reproduce 
the continuum and the overall spectral shape. We refer to Jiang et al. for any
discussion about the spectral decomposition and the study of the soft excess and
the reflection as well as their trends with the flux.
We just notice that the primary, power-law, component progressively softens
with increasing flux, whilst the blackbody temperature increases.
The energy of the Fe L line significantly increases with the flux
{most likely due to an increase of the disc ionisation at high fluxes}.
Unfortunately, the spectra are not sensitive enough to constrain any trend in the Fe K line
with the 10 spectra.

\begin{figure}
  \includegraphics[width=0.8\columnwidth,angle=0]{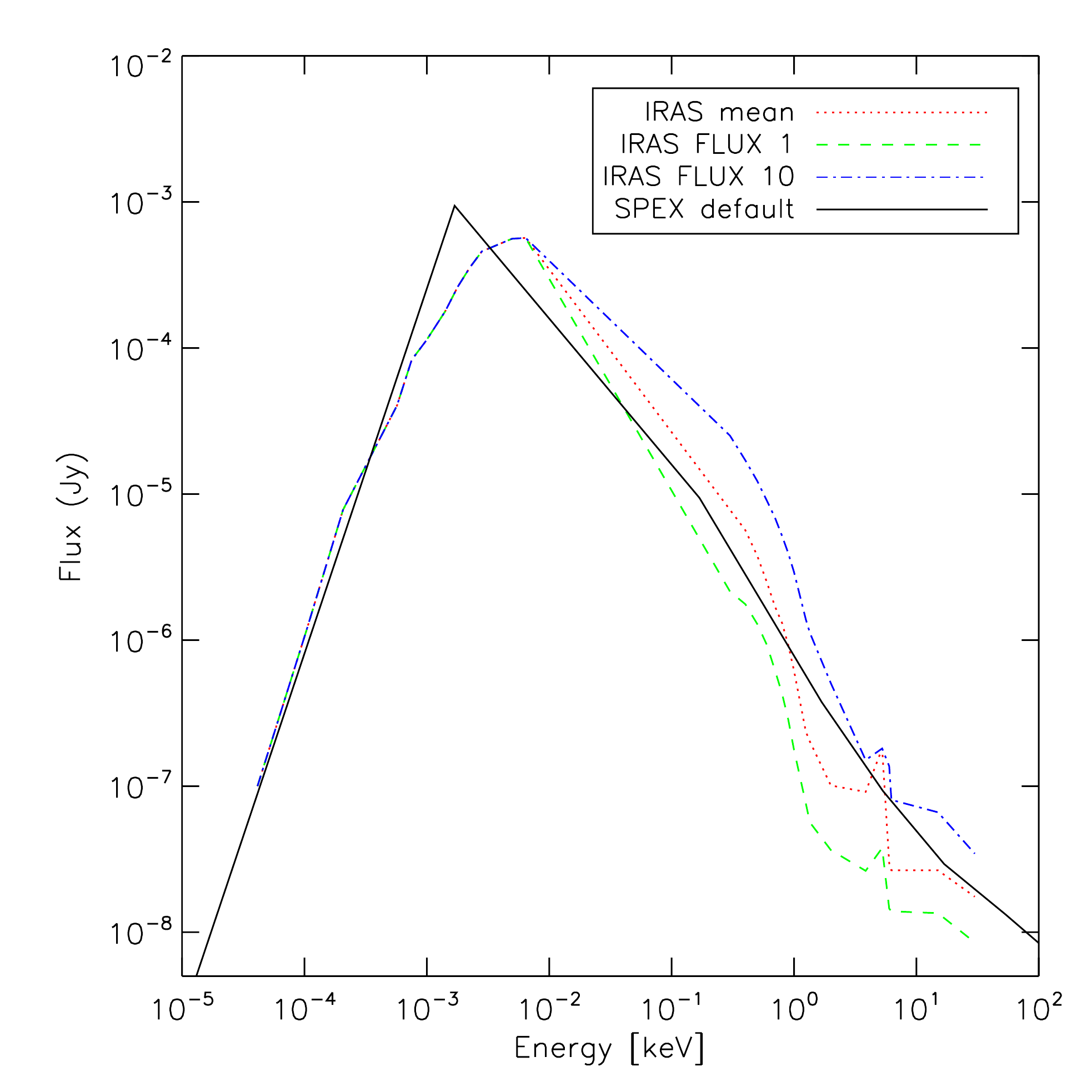}
  \centering
   \caption{Spectral Energy Distribution (SED) of IRAS 13224 lowest and highest
                flux intervals compared with the time average SED and the default SED 
                used by {\scriptsize{SPEX}} (NGC 5548).} 
            \label{Fig:Fig_SED}
     \vspace{-0.5cm}       
\end{figure}

\begin{figure}
  \includegraphics[width=0.75\columnwidth,angle=0]{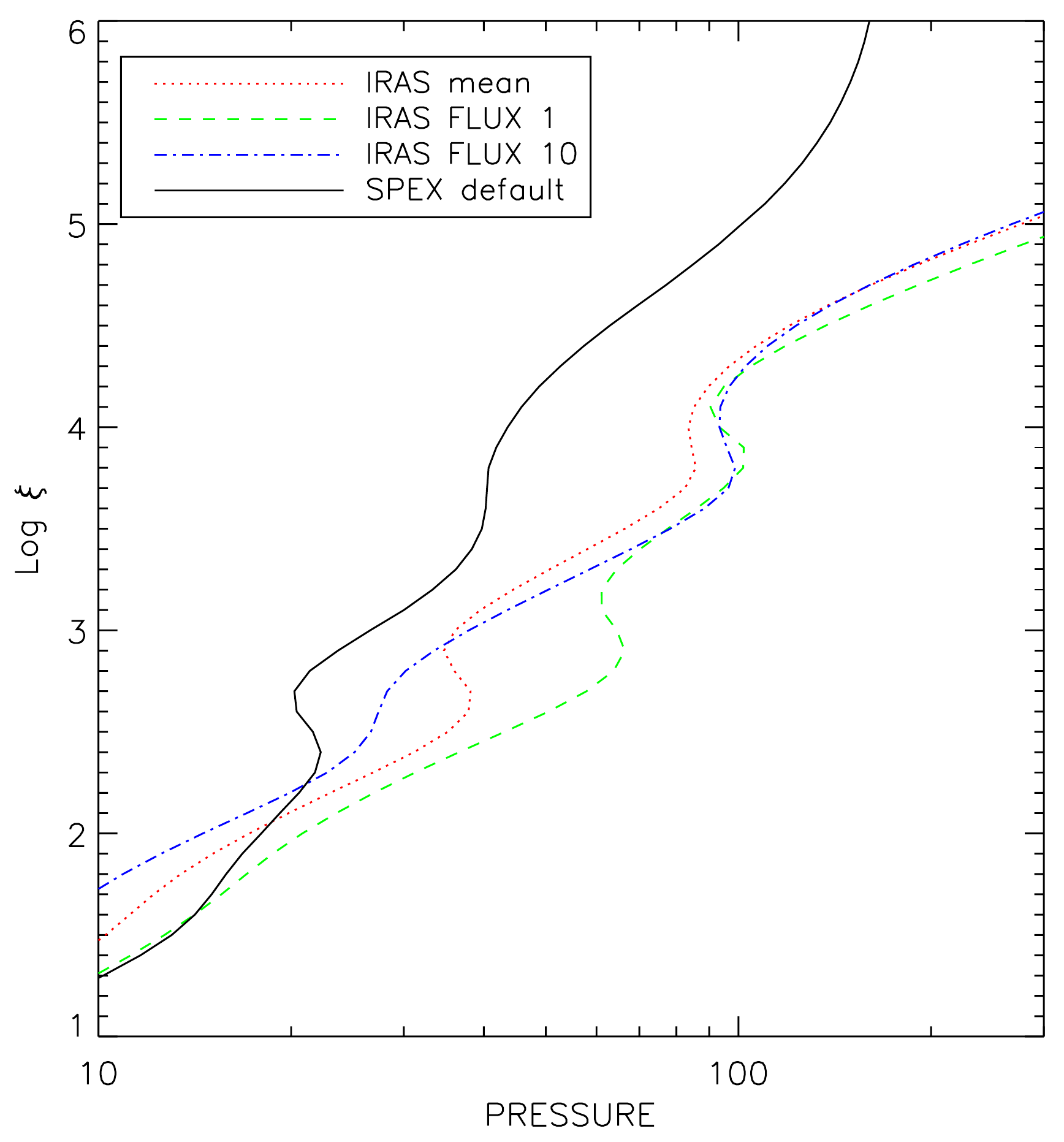}
  \centering
   \caption{Stability curves of IRAS 13224 lowest and highest
                flux intervals compared with those for the time average SED and the default SED 
                used by {\scriptsize{SPEX}} (NGC 5548, see Fig. \ref{Fig:Fig_SED}).
                The Pressure is defined as $\Xi = 19222 \, \xi / T$ (Mehdipour et al. 2016).} 
            \label{Fig:Fig_stability_curves}
     \vspace{-0.5cm}       
\end{figure}

{In Figs. \ref{Fig:Fig_SED} and \ref{Fig:Fig_stability_curves} we compare the {\scriptsize{SPEX}} 
default spectral energy distribution (SED) with the time-integrated average SED of IRAS 13224 
and those for the flux intervals 1 and 10  (which are expected to be the most different ones). 
The stability curves are calculated with the {\scriptsize{SPEX XABSINPUT}} tool. 
There is very little deviation between SED and ionisation balance calculated for different flux intervals. 
Moreover, the IRAS flux-resolved SEDs are averaged throughout several epochs and therefore 
do not take into account epoch-by-epoch variations. 
For these reasons, we preferred to use the {\scriptsize{SPEX}} default SED and ionisation balance in our fits,
just like previously done in Parker et al. 2017.
The main (systematic) effect using {\scriptsize{SPEX}} rather IRAS 13224 (putative) flux-resolved SED
is a shift of $+0.5$ in the $\log \xi$ of all our fits (as mentioned in Parker et al. 2017). 
All trends are instead confirmed and at the same significance.
We aim to obtain more detail on epoch-by-epoch effects in a forthcoming paper where we focus on 
time-resolved spectroscopy and time-resolved SED modelling.}




 
 \begin{table*}
 \caption{Continuum spectral fits for the 10 flux-resolved EPIC-pn spectra}  
 \vspace{-0.2cm}
 \label{table:table_continuum}      
 \renewcommand{\arraystretch}{1.}
  \small\addtolength{\tabcolsep}{-2.5pt}
  
 \scalebox{1}{%
 \begin{tabular}{c c c c c c c c c c}     
 \hline  
                 &  \multicolumn{2}{c}{Power law} & \multicolumn{2}{c}{Blackbody}  & \multicolumn{2}{c}{Fe L line} &  \multicolumn{2}{c}{Fe K line}     \\
                 
 Level       &  Norm  & $\Gamma$  &  Area & k$T$ & Norm & Energy & Norm & Energy \\
 
        &  $10^{52}$ ph/s/keV  &    &  $10^{23}$ cm$^{2}$ & keV & $10^{51}$ ph/s & keV & $10^{50}$ ph/s & keV \\
 
 \hline                                                                                   
                                                           
 1 & $ 0.087 \pm 0.007 $ & $ 2.189 \pm 0.009 $ & $ 1.56 \pm 0.02 $ & $ 0.1117 \pm 0.0002 $ & $ 1.24 \pm 0.02 $ & $ 0.872 \pm 0.002 $ & $ 0.76 \pm 0.03 $ & $ 7.37 \pm 0.4 $ \\
                                                         
 2 & $ 0.239 \pm 0.002 $ & $ 2.348 \pm 0.008 $ & $ 3.31 \pm 0.07 $ & $ 0.1099 \pm 0.0005 $ & $ 3.01 \pm 0.03 $ & $ 0.892 \pm 0.002 $ & $ 0.94 \pm 0.05 $ & 7.37 coupled \\
                                                         
 3 & $ 0.360 \pm 0.002 $ & $ 2.438 \pm 0.008 $ & $ 3.87 \pm 0.02 $ & $ 0.1102 \pm 0.0006 $ & $ 3.57 \pm 0.04 $ & $ 0.908 \pm 0.002 $ & $ 0.99 \pm 0.06 $ & 7.37 coupled  \\
                                                         
 4 & $ 0.532 \pm 0.003 $ & $ 2.563 \pm 0.008 $ & $ 4.09 \pm 0.09 $ & $ 0.1127 \pm 0.0006 $ & $ 4.02 \pm 0.05 $ & $ 0.916 \pm 0.003 $ & $ 1.27 \pm 0.07 $ & 7.37 coupled  \\
                                                         
 5 & $ 0.741 \pm 0.004 $ & $ 2.673 \pm 0.008 $ & $ 4.4 \pm 0.1 $ & $ 0.1125 \pm 0.0003 $ & $ 4.74 \pm 0.06 $ & $ 0.921 \pm 0.003 $ & $ 1.5 \pm 0.1 $ & 7.37 coupled  \\
                                                         
 6 & $ 0.883 \pm 0.004 $ & $ 2.690 \pm 0.007 $ & $ 4.3 \pm 0.1 $ & $ 0.1153 \pm 0.0004 $ & $ 4.6 \pm 0.1 $ & $ 0.935 \pm 0.004 $ & $ 1.5 \pm 0.1 $ & 7.37 coupled  \\
                                                         
 7 & $ 1.081 \pm 0.006 $ & $ 2.807 \pm 0.008 $ & $ 4.6 \pm 0.2 $ & $ 0.115 \pm 0.001 $ & $ 5.3 \pm 0.1 $ & $ 0.937 \pm 0.005 $ & $ 1.8 \pm 0.1 $ & 7.37 coupled  \\
                                                         
 8 & $ 1.404 \pm 0.007 $ & $ 2.792 \pm 0.007 $ & $ 3.9 \pm 0.1 $ & $ 0.120 \pm 0.001 $ & $ 4.3 \pm 0.1 $ & $ 0.967 \pm 0.004 $ & $ 1.5 \pm 0.1 $ & 7.37 coupled  \\
                                                         
 9 & $ 2.096 \pm 0.009 $ & $ 2.896 \pm 0.006 $ & $ 3.3 \pm 0.1 $ & $ 0.124 \pm 0.001 $ & $ 4.4 \pm 0.1 $ & $ 0.973 \pm 0.005 $ & $ 2.2 \pm 0.2 $ & 7.37 coupled  \\
                                                         
10 & $ 2.98 \pm 0.01 $    & $ 2.833 \pm 0.006 $ & $ 3.2 \pm 0.1 $ & $ 0.132 \pm 0.001 $ & $ 4.1 \pm 0.2 $ & $ 0.996 \pm 0.008 $& $ 1.7 \pm 0.2 $ & 7.37 coupled  \\

 \hline                
 \end{tabular}}
 
Notes: each component is redshifted ($z=0.0658$, which in {\scriptsize{SPEX}} corresponds
to a distance of 296 Mpc) and absorbed by interstellar gas. 
The two iron lines are relativistically broadened (see Sec.\,\ref{sec:appendix}).
The Fe K line energy is coupled between all models.


 \end{table*}

 \begin{table*}
 \caption{Photoionised absorber parameters for the two solutions.}  
 \vspace{-0.2cm}
 \label{table:table_absorbers}      
 \renewcommand{\arraystretch}{1.}
  \small\addtolength{\tabcolsep}{-2.5pt}
  
 \scalebox{1}{%
 \begin{tabular}{c c c c c c c c c c}     
 \hline  
                 &  & \multicolumn{3}{c}{Fast solution}  & \multicolumn{3}{c}{Slow solution}     \\
                 
 Level       &  $L_X$  & $N_{\rm H}$  &  $\log \xi$ & $v$ & $N_{\rm H}$  &  $\log \xi$ & $v$  \\
 
           &  $10^{43}$ erg s$^{-1}$ &  $10^{24}$ cm$^{-2}$  &  $\log$ (erg cm s$^{-1}$) &  $c$   & $10^{24}$ cm$^{-2}$  & $\log$ (erg cm s$^{-1}$) &  $c$ \\
 
 \hline                                                                                   
                                                           
   1 & $ 2.45 $ & $ 0.8 \pm 0.2 $ & $ 4.20 \pm 0.04 $ & $ 0.238 \pm 0.005 $ & $ 7.5 \pm 0.3 $ & $ 4.75 \pm 0.04 $ & $ 0.212 \pm 0.007 $ \\
                                                                                         
   2 & $ 5.62 $ & $ 0.8 \pm 0.2 $ & $ 4.16 \pm 0.04 $ & $ 0.238 \pm 0.005 $ & $ 7.2 \pm 0.5 $ & $ 4.74 \pm 0.04 $ & $ 0.210 \pm 0.007 $ \\
                                                                                         
   3 & $ 7.00 $ & $ 0.6 \pm 0.2 $ & $ 4.24 \pm 0.04 $ & $ 0.250 \pm 0.006 $ & $ 7.2 \pm 0.6 $ & $ 4.78 \pm 0.04 $ & $ 0.215 \pm 0.008 $ \\
                                                                                         
   4 & $ 8.42 $ & $ 0.8 \pm 0.1 $ & $ 4.20 \pm 0.04 $ & $ 0.238 \pm 0.005 $ & $ 6.9 \pm 0.4 $ & $ 4.77 \pm 0.03 $ & $ 0.211 \pm 0.007 $ \\
                                                                                         
   5 & $ 10.3 $ & $ 0.5 \pm 0.2 $ & $ 4.26 \pm 0.05 $ & $ 0.255 \pm 0.008 $ & $ 5.8 \pm 0.3 $ & $ 4.79 \pm 0.04 $ & $ 0.221 \pm 0.009 $ \\
                                                                                         
   6 & $ 11.2 $ & $ 0.6 \pm 0.2 $ & $ 4.31 \pm 0.06 $ & $ 0.247 \pm 0.007 $ & $ 6.7 \pm 1.2 $ & $ 4.86 \pm 0.04 $ & $ 0.22 \pm 0.01 $ \\
                                                                                         
   7 & $ 13.0 $ & $ 0.5 \pm 0.2 $ & $ 4.34 \pm 0.07 $ & $ 0.25 \pm 0.01 $ & $ 6.6 \pm 0.5 $ & $ 4.88 \pm 0.04 $ & $ 0.22 \pm 0.01 $ \\
                                                                                         
   8 & $ 14.5 $ & $ 0.5 \pm 0.3 $ & $ 4.6 \pm 0.1 $ & $ 0.26 \pm 0.01 $ & $ 6.1 \pm 1.7 $ & $ 5.03 \pm 0.06 $ & $ 0.23 \pm 0.01 $ \\
                                                                                         
   9 & $ 18.1 $ & $ 0.8 \pm 0.5 $ & $ 4.5 \pm 0.1 $ & $ 0.27 \pm 0.01 $ & $ 5.9 \pm 1.9 $ & $ 5.07 \pm 0.07 $ & $ 0.24 \pm 0.02 $ \\
                                                                                         
 10 & $ 24.2 $ & $ 0.5 \pm 0.3 $ & $ 4.8 \pm 0.2 $ & $ 0.29 \pm 0.03 $ & $ 8.7 \pm 1.2 $ & $ 5.7 \pm 0.3 $ & $ 0.26 \pm 0.02 $ \\

 \hline                
 \end{tabular}}
 
Notes: X-ray luminosities (0.3-10 keV) are corrected for Galactic absorption and redshift ($z=0.0658$), 
which in {\scriptsize{SPEX}} corresponds 
to a distance of 296 Mpc). The spectral fits are also shown in Fig.\,\ref{Fig:Fig_pn_rgs_spectra} and \ref{Fig:Fig_pn_rgs_residuals}.
Line of sight velocities are in units of speed of light.


 \end{table*}

\begin{figure}
  \includegraphics[width=1.02\columnwidth,angle=0]{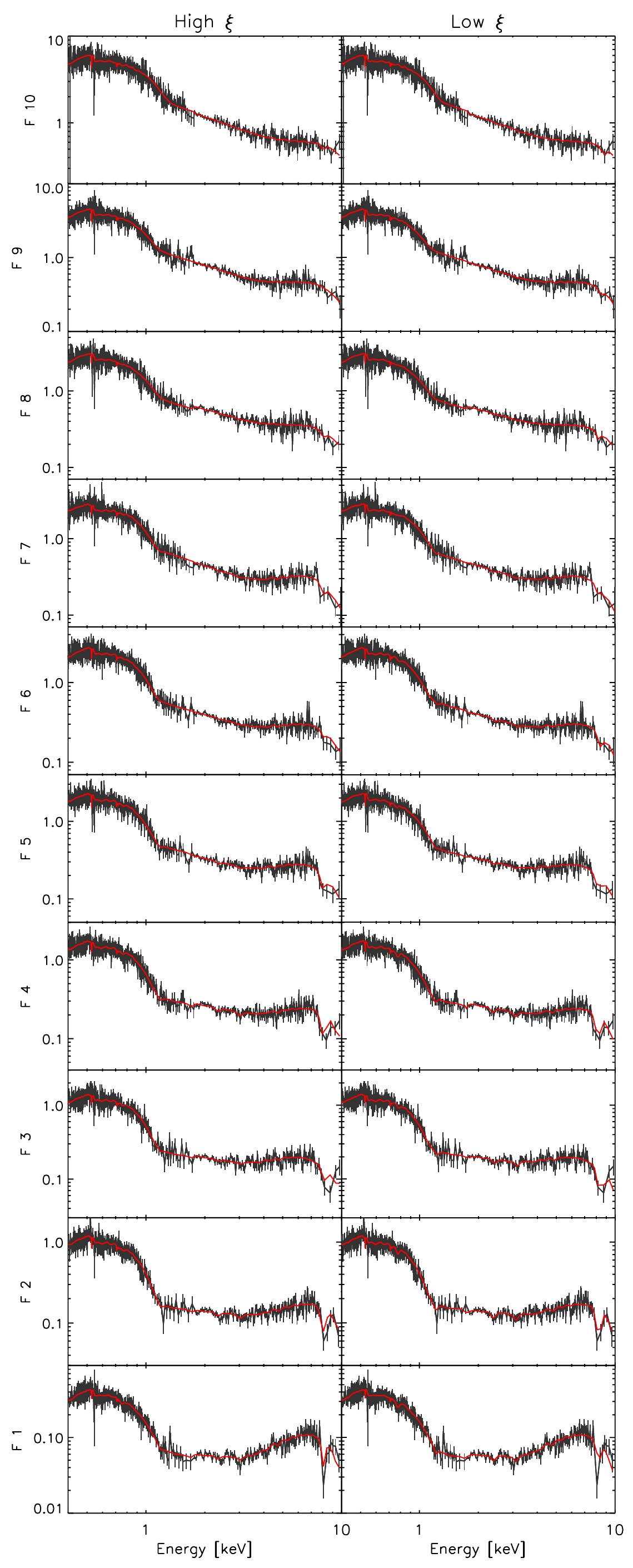}
  \centering
   \caption{IRAS 13224 flux-selected RGS ($<1.77$\,keV) and EPIC/pn ($>1.77$\,keV) spectra 
                (ordered according to increasing luminosity from bottom to top)
                with best fit models for the two solutions.
                Y-units are in counts m$^{-2}$ s$^{-1}$  \,{\AA}.} 
            \label{Fig:Fig_pn_rgs_spectra}
     \vspace{-0.5cm}       
\end{figure}

\begin{figure}
  \includegraphics[width=1.02\columnwidth,angle=0]{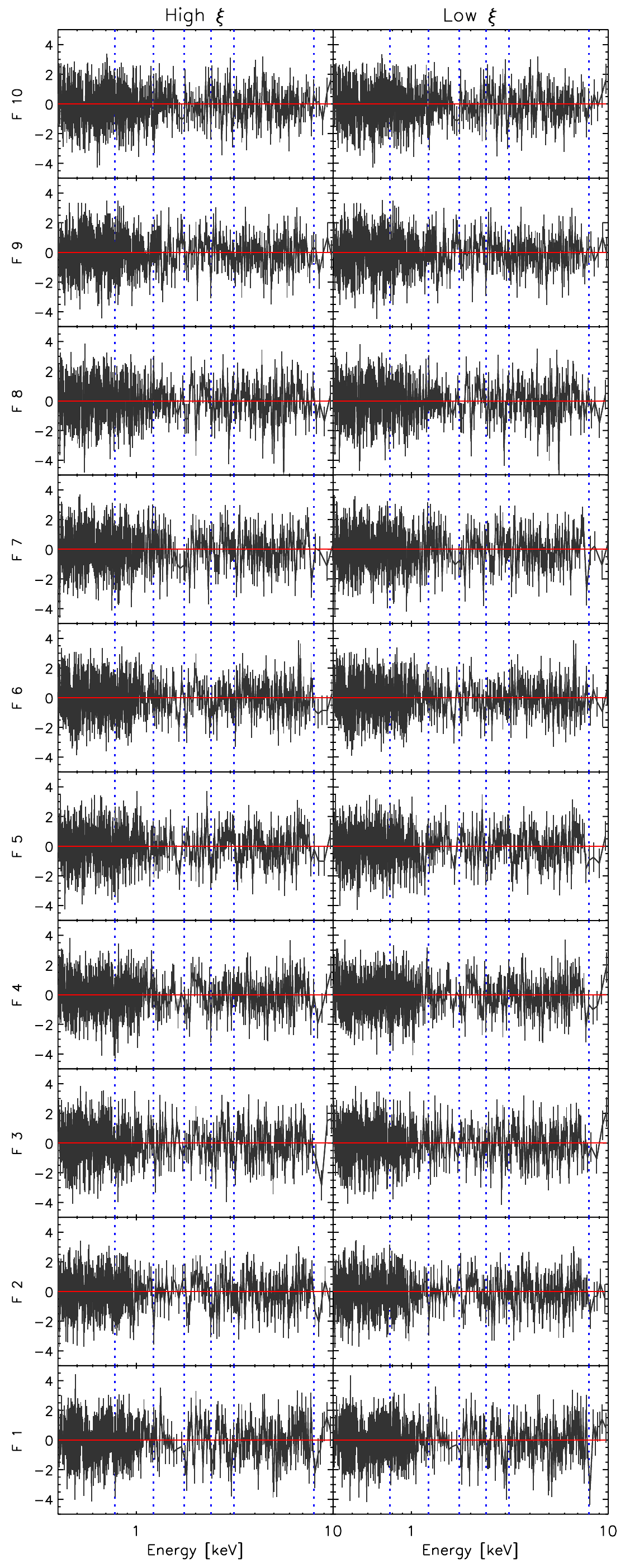}
  \centering
   \caption{IRAS 13224 flux-selected RGS ($<1.77$\,keV) and EPIC/pn ($>1.77$\,keV) spectral 
                residuals from the best fit models (see also Fig.\,\ref{Fig:Fig_pn_rgs_spectra}).} 
            \label{Fig:Fig_pn_rgs_residuals}
     \vspace{-0.5cm}       
\end{figure}


\begin{figure}
  \includegraphics[width=1.02\columnwidth,angle=0]{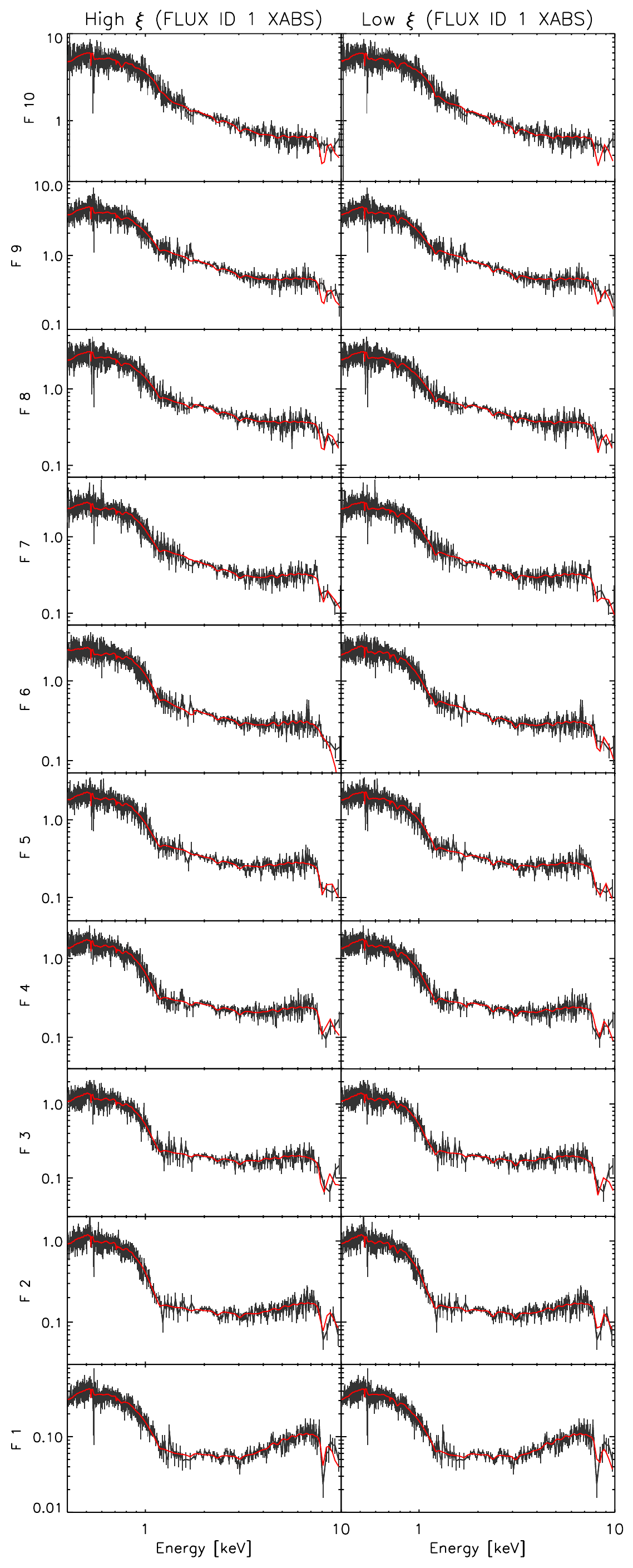}
  \centering
   \caption{IRAS 13224 flux-selected RGS ($<1.77$\,keV) EPIC/pn ($>1.77$\,keV) spectra 
                (ordered according to increasing luminosity from bottom to top)
                with absorption model from the lowest flux spectrum.
                Y-units are in counts m$^{-2}$ s$^{-1}$  \,{\AA}.} 
            \label{Fig:Fig_pn_rgs_spectra_model1}
     \vspace{-0.5cm}       
\end{figure}

\begin{figure}
  \includegraphics[width=1.02\columnwidth,angle=0]{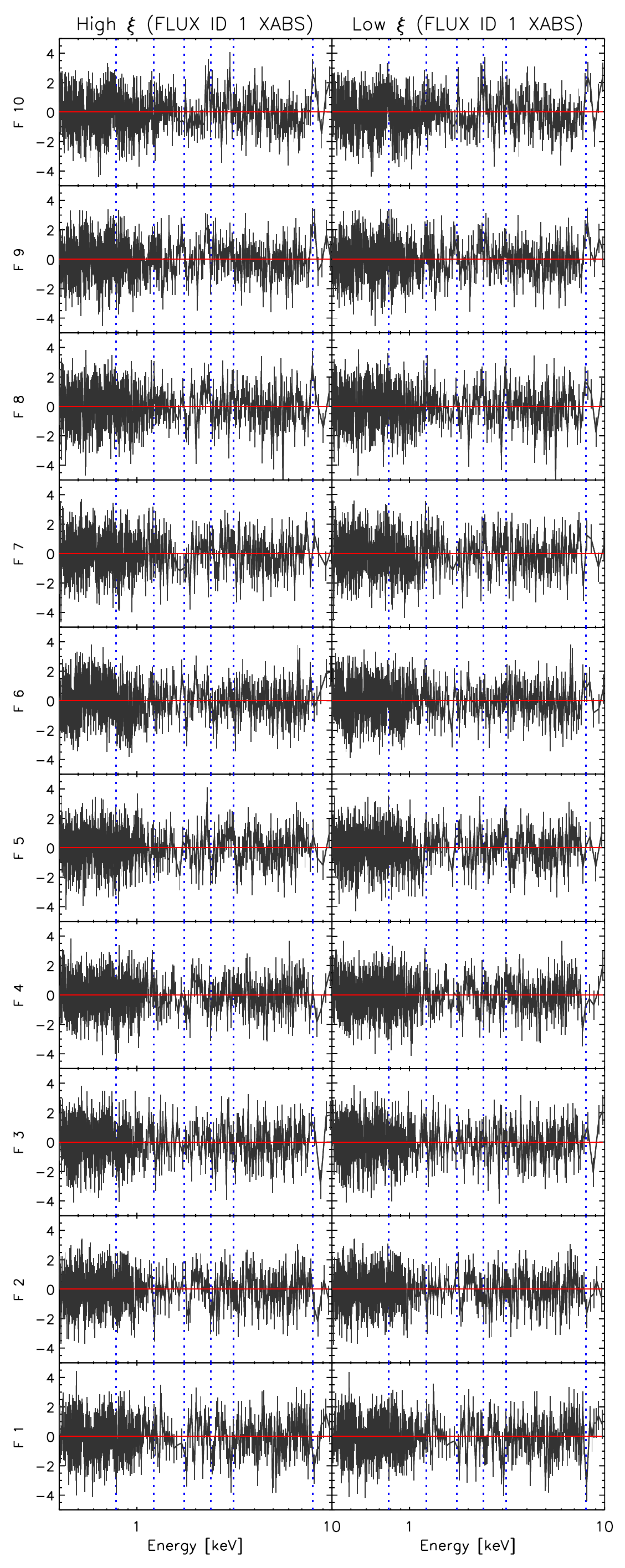}
  \centering
   \caption{IRAS 13224 flux-selected RGS ($<1.77$\,keV) EPIC/pn ($>1.77$\,keV) spectral 
                residuals adopting the absorption model parameters from the lowest flux spectrum
                (see Fig.\,\ref{Fig:Fig_pn_rgs_spectra_model1}).} 
            \label{Fig:Fig_pn_rgs_residuals_model1}
     \vspace{-0.5cm}       
\end{figure}

\label{lastpage}

\end{document}